\documentclass[12pt]{iopart}
\usepackage{graphicx,iopams}
\usepackage{esint} 
\begin{document} 
\title[Formfactor perturbation expansions and confinement in IFT]
{Formfactor perturbation expansions and confinement in the Ising field theory}
\author{S.~B. Rutkevich}
\address{Institute of Solid State and Semiconductor Physics,  SSPA  
"Scientific-Practical Materials Research Centre, 
NAS of Belarus", P.~Brovka St. 17, 220072 Minsk, Belarus}
\begin{abstract}
We study the particle spectrum $M_n(h)$ in the two-dimensional ferromagnetic Ising field theory in a 
weak external magnetic field $h$. According to Wu and McCoy scenario  
of the weak confinement, pairs of fermions (domain walls) are
coupled into bosonic kink-antikink bound states at small $h>0$. 
Fluctuations with more than two fermions also contribute to  
the wave functions of the compound particles, leading to the 
multi-fermion corrections to their masses $M_n(h)$ in higher orders in $h$. We describe a perturbative 
procedure, which allows to account both  multi-fermion fluctuations, and the 
long-range confining interaction between fermions, and leads to the formfactor expansions
for the renormalized parameters of the model.  We obtain 
integral representations for the third-order multi-fermion correction to the mass 
$M_n(h)$, which arise from the regular correction to the  kernel of the Bethe-Salpeter equation. 		
\end{abstract}
\ead{rut@ifttp.bas-net.by}
\maketitle
\section{Introduction}
In recent years much progress has been achieved in the understanding of the scaling limit of the  two-dimensional
Ising model, which is known  as the  Ising Field Theory (IFT), for a review see \cite{Del04}.
Providing direct information about the Ising universality class in two dimensions, IFT can be viewed
also as a continuous dynamical model of the one-dimensional uniaxial ferromagnet. Being, perhaps, 
the simplest relativistic model describing 
confinement of topological excitations, IFT can give a deep insight into some
nontrivial aspects of  confinement in the particle and condense matter physics. 

IFT contains parameters $m$ and $h$, which are proportional to the deviations of temperature 
$T$ and magnetic field $H$ from their critical values in the two-dimensional lattice Ising model,
$m\sim (T_c-T)$, $h\sim H$. At the critical point $m=0$, $h=0$, IFT 
reduces to the Conformal Field Theory with central charge $c=\frac{1}{2}$, 
which Euclidean action $\mathcal{A}_{CFT}$ describes free massless Majorana fermions. 
It has two relevant operators, the energy density $\varepsilon(x)$, and the order
spin operator $\sigma(x)$. IFT can be defined as the perturbation of Ising Conformal Field Theory
by these two  operators, which is described by the action  \cite{FZ06}
\begin{equation}
{\mathcal A}_{IFT}={\mathcal A}_{CFT}+2\pi m \int\varepsilon(x)\, d^2 x-h \int\sigma(x)\, d^2 x.
\label{AIFT0}
\end{equation}
 In fact, only one dimensionless parameter
$\eta=m/|h|^{8/15}$ determines the physics of IFT.

IFT being not integrable for generic $h$ and $m$, admits exact solutions 
along the directions $h=0$ and $m=0$.  
The line $h=0$, $m \ne0$
corresponds to Onsager's   exact solution \cite{Ons44}. Fermions remain free here, 
but gain the mass $|m|$. In the disordered (paramagnetic) 
phase $m<0$ these fermions
are ordinary particles, while in the ordered (ferromagnetic) phase $m>0$ they are interpreted as 
topological excitations (kinks), 
which separate regions with oppositely directed spontaneous magnetization. 
Nonzero magnetic field $h>0$ induces interaction
between fermions, breaking integrability of IFT at $m\ne0$. On the other hand,
IFT has a remarkable exact solution at $m=0$,  $h\ne0$ containing 
eight  massive particles, which was found by  A.~B.~Zamolodchikov \cite{ZamH}. 

Beyond the integrable directions, IFT can be  studied by  approximate
methods - numerical and analytical. An effective numerical method 
known as Truncated Conformal Space Approach was invented by
Yurov and Alexei B. Zamolodchikov \cite{YuZam90}, \cite{YuZam91}. 
Fonseca and A.~B. Zamolodchikov \cite{FonZam2003} modified this technique and applied it 
to analysis of analytical properties of the IFT free
energy continued to complex values of the scaling parameter $\eta$. 

For analytical study of IFT for $h$ and $m$ close to the integrable directions,
it is natural to exploit perturbation expansions. Form-factor perturbation theory developed by
Delfino, Mussardo and Simonetti \cite{Del96} has been applied \cite{Del96}, \cite{DGM06} 
to calculate the   variation of the
particle mass spectrum and the decay widths of non-stable particle for small $\eta$, i.e. near the line
$m=0$. 
One could expect, that the perturbation expansion at $m\ne 0$ and small $h$ should be more simple, 
since (unperturbed) IFT is free at $h=0$. Though this is really the case in the high-temperature phase $m<0$, the
small-$h$ expansion at $m>0$ turns out to be rather non-trivial due to
the long-range attractive potential between neighbouring fermions, which is induced by the external magnetic field $h>0$.
This attractive interaction can not be accounted by the straightforward formfactor perturbative theory
at small values of $h$, and leads to
confinement of fermions. 

\begin{figure}[htb]
\centering
\includegraphics[width= .8\linewidth]{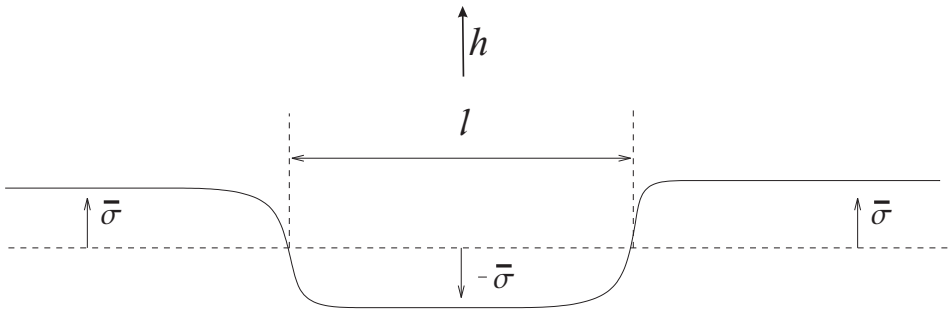}
\caption{Two kinks interact with the energy $2h\bar{\sigma}l$.\label{figure1}} 
\end{figure}

The effect of a small magnetic field $h$, which brakes the $\mathbb{Z}_2$-symmetry in the ordered phase $m>0$ in IFT,
 can be qualitatively understood by
the following simple arguments first developed by McCoy and Wu \cite {McCoy78}. 
At $h=0$, two ferromagnetic ground states $\mid 0_{+}\rangle $ and $\mid 0_{-}\rangle $ with spontaneous magnetizations $+\bar{\sigma}$
and $-\bar{\sigma}$ have the same energy. 
A weak magnetic field $h>0$ removes degeneration decreasing the energy of the state $\mid 0_{+}\rangle $,
and increasing the energy of the state $\mid 0_{-}\rangle $, which becomes metastable. In order to generate  a domain
of the metastable phase in the stable surrounding, one needs to add the energy proportional to the length of the domain.
In other words, two domain walls bounding such a domain attract one another with the energy $2 h\, \bar{\sigma}\, l$ 
proportional to their separation $l$, see figure~\ref{figure1}.
The long-range attraction leads to confinement: all domain walls are coupled into pairs 
at arbitrary small $h>0$.
Elementary excitations now are the domains bounded by two kinks, while an isolated kink gains infinite energy. 

The mechanism of confinement outlined above is quite general in one-dimensional systems. 
It is realized in such continuous one-dimensional models 
as multi-frequency Sin-Gordon model \cite{DelMus98}, $q$-state Potts field theory \cite{DG08}, and in the discrete Ising spin chain
\cite{Rut08}. Confinement of topological excitations in one-dimensional antiferromagnet has been observed 
experimentally by Kenzelmann {\it et al.} \cite{Kenz05}. On the other hand, there is a lot of similarities
between confinement in the IFT and in 't~Hooft's model for
two-dimensional multicolor QCD \cite{Hooft74}, see the discussion in \cite{FZ06}. 
Accordingly, the fermions and their bound states in IFT in the confinement regime 
are used to call as "quarks" and "mesons", respectively. 

At small $h$ the weak confinement regime is realized in IFT. In this regime, the mass spectrum $M_n$  of mesons 
is dense in the segment $[2m,\infty)$. Two  asymptotic expansions describe  $M_n$
at $h\to0$ in different regions of this segment. Near the  edge point $2m$ (i.e. for
fixed $n$ at $h\to0$)   one can use the {\it low energy expansion} in fractional powers
of the magnetic field \cite{FonZam2003}, \cite{FZ06}.
On the other hand, for $n\gg1$ and  $h\to0$,  the {\it semiclassical expansion} in integer powers
of $h$ can be applied \cite{FZ06}, \cite{Rut05}.
Derivation of the both expansions  are based on the perturbative analysis 
of the Bethe-Salpeter  equation, which determines the 
meson mass and wave function in the {\it two-quark approximation}. 
The latter implies, that one approximates the 
meson wave-function (the eigenstate of the IFT Hamiltonian) by the two-quark state, neglecting multi-quark
(four-quark, six-quark, \dots) contributions to it. The two-quark approximation is asymptotically
exact in the limit $h\to 0$  giving correct meson masses in the leading order in $h$. 
However, starting from the second order in $h$, it is necessary to take  into account the virtual multi-quark fluctuations. 
Note, that multi-quark effects are essential also for such interesting phenomena as decay of unstable mesons, and 
inelastic meson scattering. 

The second order multi-quark correction  to the meson mass was obtained by Fonseca and A.B.~Zamolodchikov
 \cite{FZ06}.  These authors demonstrated also, that
the multi-quark corrections could come up in the weak-coupling expansions of the meson masses $M_n$ in three  ways:
\begin{enumerate}
\item \label{disp}
through the radiative corrections of the quark mass and self energy;
\item \label{st}
by renormalization of the long-range attractive force between the neighbouring quarks (the "string-tension");
\item \label{pint}
by modifying the regular part of the Bethe-Salpeter kernel, which is responsible for the pair 
interaction between quarks at short distances. 
\end{enumerate}
It turns out, that only the first contribution (\ref{disp}) 
gives rise to the second-order correction to the meson mass, while (\ref{st})
and (\ref{pint}) should show up only in the third order correction, which is still unknown. 

Extension of the weak-coupling expansions for the meson masses to the third order in the magnetic field
presents an interesting problem, which we address  in this work. It could give us some insight into the role 
of the multi-particle fluctuations in the composite particles in non-integrable models exhibiting confinement.
Since multi-quark effect are responsible also for the decay 
of unstable mesons, this should manifest itself in some form in the perturbative meson mass spectrum 
near and above the stability threshold. Note, that an accurate numerical calculation of the lowest 
meson masses was reported in reference \cite{FZ06}, which clearly indicates contribution of the multi-quark 
fluctuations. 
 
Since the problem outlined above  is rather involved, here we shall
concentrate only on three parts of it. First, we extend the semiclassical  expansion 
of the original (written in the two-quark approximation) Bethe-Salpeter equation to the third order in $h$. 
Second, describe the formfactor perturbative technique, which is suitable to deal with the  multi-particle 
fluctuations in systems with confinement. Finally, we obtain the integral representations for the "local" multi-quark
correction of the meson masses, i.e. corrections (\ref{pint}) induced by renormalization of the {\it local} 
interaction between quarks.

The paper is organized as follows. In section~\ref{Mod} we describe definition of IFT and 
its operator content.  In two subsequent  sections we summarize briefly the recent progress in the theory 
of the weak confinement in  IFT: section~\ref{SBS} introduces the Bethe-Salpeter equation and its
 weak-coupling expansions, and  section~\ref{MQ} contains preliminary discussion of the multi-quark corrections
to the meson masses. 
In section~\ref{DIAH} we develop a formfactor perturbative procedure, which is modified to 
a system with a long-range confining interaction between fermions.  
It is based on the partial diagonalization of the Hamiltonian in the fermionic number, and  allows one to 
effectively account the multi-quark fluctuations by   "dressing"  the fermionic operators.
In section~\ref{MCC} we describe a compact integral
representation for the local third order correction to the meson mass, which is analysed further in  \ref{integ}.
\ref{ExBS} contains perturbative solution of the "bare" Bethe-Salpeter equation to the third order in $h$.
Concluding remarks are presented in section~\ref{CONC}.
%%%%%%%%%%%%%%%%%%%%%%%%%%%%%%%%%%%%%%%%%%%%%%%%%%%%%%%%%%%%%%%%%%%%%%%%%%%%%%%%%%%%%%%%%%%%%%%%%%%%%%%%%%%%%%%
\section{The model \label{Mod}}
Ising field theory is the Euclidean field theory, which describes the scaling limit of the 
two-dimensional lattice Ising model in the critical region $T\to T_c$, $H\to 0$. 
It is defined by the action
\begin{equation}
 \mathcal{A}_{IFT} = \frac{1}{2 \pi}\int_{-\infty}^\infty \Big[ \psi \overline{\partial}
\psi +\overline{\psi} \partial \overline{\psi}+
\rmi\, m \overline{\psi} \psi \Big] \rmd^2 x-h \int_{-\infty}^\infty \rmd^2 x\; \sigma (x).
\label{AIFT}
\end{equation}
Here $x$ denotes a point in the plane $\mathbb{R}^2$ with cartesian coordinates 
$\langle{\rm{x}}(x),{\rm{y}}(x)\rangle$ and the complex coordinate $z={\rm x}+\rmi {\rm y},$
$ \partial=\frac{1}{2}(\partial_{\rm{x}}-\rmi\partial_{\rm{y}}),\quad
\overline{\partial}=
\frac{1}{2}(\partial_{\rm{x}}+\rmi\partial_{\rm{y}}).$
Action (\ref{AIFT}) is covariant under rotation, and becomes Lorentz covariant 
after the Wick turn ${\rm y}\to \rmi t$.

Corresponding to action (\ref{AIFT}) Hamiltonian  can be written in the form
%\numparts 
\begin{eqnarray}
&&\mathcal{H} = \mathcal{H}_0 +h\,V, \label{Ham}\\
\textrm{where   }\;\;\;&&\mathcal{H}_0=\int_{-\infty}^\infty \frac{\rmd p}{2 \pi} 
\,\omega(p)\, {\bf a}^\dagger (p) \, {\bf a}(p),  \nonumber\\
&&V=-\int_{-\infty}^\infty  \rmd\rm{x}\,\sigma(\rm{x}), \nonumber%  \label{V}
\end{eqnarray}
%\endnumparts \label{77}
and  
$\omega(p)=(p^2+m^2)^{1/2}$ is the spectrum of free fermions. 
Fermionic operators $ {\bf a}^\dagger (p') , \,{\bf a}(p)$  obey the canonical 
anticommutational relations
\[
\{ {\bf a}(p) , {\bf a}^\dagger (p') \} =2 \pi \,\delta(p-p'), \quad 
 \{ {\bf a}(p), {\bf a} (p') \} = \{ {\bf a}^\dagger (p) ,{\bf a}^\dagger (p') \}= 0.  
\]
Commonly used are also fermionic operators $a(\beta),\, a^\dagger(\beta)$, 
corresponding to the rapidity variable $\beta={\rm arcsinh}(p/m)$:
\[
a(\beta)= \omega(p)^{1/2}\, {\bf a}(p), \;a^\dagger(\beta)=\omega(p)^{1/2}\, {\bf a}^\dagger (p). 
\]
Notations 
\begin{eqnarray*}
|p_1, \dots ,p_N \rangle = {\bf a}^\dagger (p_1) \dots  {\bf a}^\dagger (p_N) |0\rangle, \quad\;\;
\langle p_1,\ldots,p_N|=\langle 0|{\bf  a}(p_1) \dots  {\bf  a}(p_N) , \\
| \beta_1,\dots, \beta_N \rangle = a^\dagger(\beta_1) \dots a^\dagger(\beta_N|0\rangle, \quad\quad
\langle \beta_1,\dots ,\beta_N|=\langle 0|  a(\beta_1) \dots a(\beta_N)
\end{eqnarray*}
for the fermionic basis states with definite momenta will be used. 

The order spin operator $ \sigma({\rm x}) =\sigma( {\rm x} ,{\rm y})|_{{\rm y}=0}$ in 
the ordered phase $T<T_C$  (i.e. at $m>0$)
can be determined in the  infinite line 
${\rm x}\in \mathbb{R}$ as the normally ordered 
exponent~\cite{Jimb80,Rut01}:
\begin{eqnarray}
 \sigma ({\rm x})&=&\bar{\sigma} : \rme^{\rho({\rm x)}/2} :\;, \label{sigma}\\
\frac{ \rho({\rm x})}{2}&=&  \int_{\rm x}^\infty \rmd{\rm x}' \big(\chi({\rm x}',{\rm y})\,
\partial_{\rm y} \chi ({\rm x}',{\rm y})\big){\big |}_{{\rm y}=0}, \nonumber   \\
\chi({\rm x},{\rm y})=&\rmi& \int_{-\infty}^\infty \frac{\rmd p}{2 \pi} 
\frac{\rme^{i p x}}{\sqrt{\omega(p)}}\bigg({\bf a}^{\dagger}(-p) \,
\rme^{ \omega(p)\,{\rm y}}-{\bf a}(p)\, \rme^{- \omega(p)\,{\rm y}}\bigg),  \nonumber
\end{eqnarray}
where $\bar{\sigma}=m^{1/8}2^{1/12}e^{-1/8}A^{3/2}$ is the zero-field vacuum 
expectation value of the order field (spontaneous magnetization), 
$A=1.28243...$ is Glaisher's constant.  

Alternatively, operators $\sigma({\rm x})$ can be completely characterized by their 
formfactors $ \langle \beta_1,\ldots,\beta_K|\sigma(0) |\beta'_1,\ldots,\beta'_{N}\rangle $, which 
explicit expressions are well known  
\cite{Berg79,FonZam2003}. In the ordered phase 
\begin{eqnarray} 
\langle \beta_1,\ldots,\beta_K|\sigma(0) |\beta'_1,\ldots,\beta'_{N}\rangle   \label{for}\\
\fl =\rmi^{(K+N)/2}\,\bar{\sigma} \prod_{0<i<j\le K}\tanh\left(\frac{\beta_i-\beta_j}{2}\right)
 \prod_{0<k<q\le N}\tanh\left(\frac{\beta'_k-\beta'_q}{2}\right)
\prod_{\begin{array}{ccc}{\scriptstyle 0<s\le K} \\
{\scriptstyle 0<t\le N}\end{array}} \coth\left(\frac{\beta_s-\beta'_t}{2}\right) ,       \nonumber
\end{eqnarray}
if $(K+N)$ is even, and 
$
\langle \beta_1,\ldots,\beta_K|\sigma(0) |\beta'_1,\ldots,\beta'_{N}\rangle=0
$
for odd $(K+N)$. 
The right-hand side in (\ref{for}) contains factors $\coth\left(\frac{\beta_s-\beta'_t}{2}\right)$, 
which are singular at
at $\beta_s=\beta'_t$. 
These kinematic singularities should be understood in the 
sense of the Cauchy principal value
\[ \coth\left(\frac{\beta_s-\beta'_t}{2}\right) \to
\frac{1}{4}\bigg[ \coth\left(\frac{\beta_s-\beta'_t+ \rmi 0 }{2}\right) +
 \coth\left(\frac{\beta_s-\beta'_t - \rmi 0 }{2}\right)\bigg].
\] 
Note, that the Wick expansion holds for  formfactors (\ref{for}) of the 
spin operator. 
For example, 
\begin{eqnarray*}
 \bar{\sigma} \,\langle \beta_1,\beta_2|\sigma(0) |\beta'_1,\beta'_{2}\rangle =
\langle\beta_1|\sigma(0) |\beta'_{2}\rangle \langle\beta_2|\sigma(0) |\beta'_{1}\rangle  \\
-\langle \beta_1|\sigma(0) |\beta'_{1}\rangle \langle\beta_2|\sigma(0) |\beta'_{2}\rangle
+\langle \beta_1,\beta_2|\sigma(0) |0\rangle  \langle  0|\sigma(0) |\beta'_{1},\beta'_2\rangle. 
\end{eqnarray*}
%%%%%%%%%%%%%%%%%%%%%%%%%%%%%%%%%%%%%%%%%%%%%%%%%%%%%%%%%%%%%%%%%%%%%%%%%%%%%%%%%%%%%%%%%%%
\section{Bethe-Salpeter equation \label{SBS}}
The meson energy spectra $\Delta E_n(P) $ can be formally determined from the eigenvalue problem:
\begin{eqnarray} \label{eig}
\mathcal{H} \mid \Phi _n(P)\rangle =[ \Delta E_n(P)+E_{\rm vac}]\mid \Phi _n(P)\rangle, \\
\mathcal{H}=\mathcal{H}_0+h\,V,  \nonumber \\
\hat{P}\mid \Phi _n(P)\rangle=P \mid \Phi _n(P)\rangle, \label{mom}
\end{eqnarray}
where $\hat{P}$ is the total momentum operator, 
\[
\hat{P}=\int_{-\infty}^\infty \frac{\rmd p}{2 \pi}\, p\, {\bf a}^\dagger (p) \, {\bf a}(p),
\]
and $E_{\rm vac} $ is the ground state energy, which is proportional to the length of the system $L$. 

The eigenvalue problem (\ref{eig}) is quite difficult, since the Hamiltonian contains the order spin 
operator $\sigma({\rm x})$, which is highly nonlinear in fermionic fields. A significant simplification 
can be provided by the two-quark approximation \cite{FonZam2003}, \cite{Rut99}. 
 It implies that one replaces the
exact Hamiltonian eigenvalue problem (\ref{eig}), (\ref{mom}) 
by its projection to the two-quark subspace ${\mathbb F}_2$ of the Fock space ${\mathbb F}$:
\begin{eqnarray} \label{eig2}
\mathcal{P}_2 \,\mathcal{H} \mid \widetilde{\Phi} _n(p)\rangle 
=[\Delta \widetilde{E}_n(P)+\widetilde{E}_{vac}] \mid \widetilde{\Phi}_n(P)\rangle ,\\
\hat{P}\mid \widetilde{\Phi} _n(P)\rangle=P \mid \widetilde{\Phi} _n(P)\rangle,  \nonumber\\
\mid \widetilde{\Phi}_n(P)\rangle \in \mathbb{F}_2. \nonumber
\end{eqnarray}
Here ${\mathcal P}_n$ denotes the orthogonal projector onto the $n$-quark subspace ${\mathbb F}_n$ of ${\mathbb F}$.  
Tildes distinguish solutions of (\ref{eig2})
from those of the exact eigenvalue problem (\ref{eig}).

In the momentum representation, equation (\ref{eig2}) takes the form \cite{FZ06}
\begin{eqnarray} \label{BSP}
\fl \left[ \omega(P/2+p) +\omega(P/2-p)- \Delta \widetilde{E}(P)   \right] \Psi_P(p )     
=f_0\, \fint_{-\infty}^\infty 
G_P(p|k) \, \Psi_P(k) \, \frac{\rmd k}{2 \pi},   
\end{eqnarray}
where $\fint$ denotes the Cauchy principal value integral, 
\begin{eqnarray} \label{Psi}
\langle P'/2-p, P'/2+p|\widetilde{\Phi}(P)\rangle=2\pi\delta(P'-P)\,\Psi_P(p), \\
 G_P(p|k)=\mathcal{G}(P/2+p,P/2-p|P/2+k,P/2-k), \nonumber\\
\fl  \mathcal{G}(p_1, p_2|k_1,k_2)=\frac{1}{4\bar{\sigma}}\langle \,p_2,p_1\,|\sigma(0)|\,k_1,k_2\,\rangle 
=\frac{1/4}{[ \omega(p_1) \omega(p_2)\omega(k_1) \omega(k_2) ]^{1/2}} \label{G0} \\
\fl  \cdot\Bigg[
\frac{  \omega(p_1) +\omega(k_2) }{p_1-k_2} \,\frac{ \omega(p_2) +\omega(k_1)}{p_2-k_1} -
\frac{\omega(p_1) +\omega(k_1)}{p_1-k_1} \,\frac{ \omega(p_2) +\omega(k_2)}{p_2-k_2}  \nonumber\\
\lo + \frac{p_1-p_2}{\omega(p_1) +\omega(p_2) }\, \frac{k_1-k_2}{\omega(k_1) +\omega(k_2) }          
 \Bigg], \nonumber
\end{eqnarray}
and $f_0=2 h\bar{\sigma}=\lambda\,m^2$ is the "bare string tension". Index $n$ is omitted in (\ref{BSP}), (\ref{Psi}).
Note, that $\Psi_P(p )$ is an odd function of $p$, and 
\[
G_P(p|k) = \frac{1}{(p-k)^2}- \frac{1}{(p+k)^2}+ G^{(reg)}_P(p|k) ,
\]
where $G^{(reg)}_P(p|k)$ is regular at real $p$ and $q$. The pole terms in $G_P(p|k)$ 
produce after the Fourier transform the long-range linear attractive potential $f_0 |{\rm x}|$ 
proportional to the distance $|{\rm x}|$ between the two quarks. The regular term 
$G^{(reg)}_P(p|k)$ is responsible for the local interaction between quarks vanishing
at the distances
$\gg m^{-1}$. 

Equation (\ref{BSP}) is   the Bethe-Salpeter equation written in a generic
momentum frame. It  simplifies in two  cases.
\begin{itemize}
\item In the frame of the centre of mass of two quarks \cite{FonZam2003}, $P=p_1+p_2=0$:
\begin{eqnarray}\label{BS}
 \left[2\,\omega (p)-\Delta \widetilde{E}(0)\right]\, \Psi_0 (p) = f_0
\fint_{-\infty}^\infty  \frac{\rmd k}{2 \pi}\,\frac{\Psi_0(k)}{ 2\,\omega(p)\omega(k) } \\ \cdot 
\Bigg[\left(\frac{ \omega(p)+\omega(k) }{p-k}\right)^2  +
\frac{ 1 }{2}\frac{p\,k}{\omega(p)\omega(k) }\Bigg]. \nonumber
\end{eqnarray}
\item In the infinite momentum frame (see Appendix A in \cite{FZ06}), $P\to \infty$: 
\begin{equation}\label{BSi}
\left[ \frac{m^2}{1-u^2}-\frac{\widetilde{M}^2}{4}  \right]\Phi(u)=f_0 \fint_{-1}^1
F(u|v) \,\Phi(v)\,\frac{\rmd v}{2\pi},
\end{equation}
where the scaled variables $u=(p_1-p_2)/P$ and   $v=(q_1-q_2)/P$ have been used, and  
\begin{eqnarray*}
F(u|v)= \left[(1-u^2)(1-v^2)\right]^{-1/2}\left[ \frac{1-u v}{(u-v)^2} -\frac{1+u v}{(u+v)^2}+\frac{u v}{4}  \right],\\
\Phi(u)= \lim_{P\to \infty} \Psi_P (P\,u).
\end{eqnarray*}
The following large-$P$ asymptotic behaviour of $ \Delta \widetilde{E}(P) $ 
 was assumed in \cite{FZ06} in deriving (\ref{BSi}) from (\ref{BSP}):
\begin{equation} \label{PInf}
\Delta \widetilde{E}(P) =|P|+\frac{ \widetilde{M}^2 }{2 |P|}+ O(|P|^{-3}).
\end{equation}
\end{itemize}

Bethe-Salpeter equation (\ref{BSP}) and its particular cases (\ref{BS}), (\ref{BSi}) are
the linear singular integral equations \cite{Mus}. Different  techniques \cite{FZ06}, \cite{FonZam2003}, \cite{Rut05} 
have been been developed for their perturbative solutions  
in the weak-coupling limit $\lambda\to 0$. Fonseca and A.B.~Zamolodchikov calculated \cite{FZ06} 
several initial terms in the low-energy expansion (for fixed $n$ and 
$\lambda\to 0$) for eigenvalues of equation (\ref{BSi})  
\begin{eqnarray}
\label{lowseries}
\fl  \frac{{\widetilde M}_n^2 } {4 m^2} - 1 = z_n \,t^2 +
{\frac{z_n^2}{5}}\,t^4 - \bigg({\frac{3\,z_n^3}{175}}+\frac{57}
{280}\bigg)\,t^6 + \bigg(\frac{23\,z_n^4}{ 7875} +
{\frac{1543\,z_n}{12600}}\bigg) \,t^8 + {\frac{13}{1120\, \pi}}\, t^9 \\
+
\left(-{\frac{1894\,z_n^5}{3031875}} - {\frac{23983\,z_n^2}{242550}} \right) t^{10} +
 {\frac{3313\,z_n}{10080\,\pi}}\,t^{11}+\ldots,   \nonumber
\end{eqnarray}
where $t=\lambda^{1/3}$, and $(-z_n)$ is the zero of the Airy function, ${\rm Ai}(-z_n)=0$. 
The leading term in the above expansion reproduces the old result of McCoy and Wu~\cite{McCoy78}. 

To the second order in $\lambda$, semiclassical expansions  (for $n\gg1$ and $\lambda\to 0$) for ${\widetilde M}^2$, and for $\Delta \widetilde{E}(0)$ 
 were found in references \cite{FZ06}, and \cite{Rut05}, respectively. We extend the former expansion to the third order in 
$\lambda$  using the technique, which was applied previously in the similar discrete-chain  problem \cite{Rut08}. 
This calculation is described in \ref{ExBS}, the result reads as
\begin{equation}
{{ {\widetilde M}_n^2 } \over{4 m^2}}=\cosh^2 \theta_n,
\label{2M}
\end{equation}
where $\theta_n$ solves equation 
\begin{equation}
\sinh 2\theta_n - 2\theta_n =2\pi\lambda (n-1/4) + 2\lambda^2 {S}_1(\theta_n)+2\lambda^3 {S}_2(\theta_n)+O(\lambda^4),
\label{dis}
\end{equation}
and
\begin{eqnarray}
\fl {S}_1(\theta_n)=\frac{1}{\sinh(2\theta_n)}\bigg( \frac{5}{24\sinh^2\theta_n} -\frac{1}{12}+
\frac{1}{4\cosh^2\theta_n}-\frac{\sinh^2\theta_n}{6}\bigg), \label{S1}\\
\fl {S}_2(\theta_n)=\frac{1}{192\pi\sinh^6(2\theta)}\{-999 \,\theta_n -3\theta_n[648\cosh(2\theta_n)+228\cosh(4\theta_n) \label{S2}
\\+56\cosh(6\theta_n)   +15\cosh(8\theta_n)]
+546\sinh(2\theta_n) +363\sinh(4\theta_n)\nonumber\\ +170\sinh(6\theta_n)+
33\sinh(8\theta_n)+\sinh(12\theta_n)\}.\nonumber
\end{eqnarray}
To the second order, (\ref{2M})-(\ref{S2}) agrees with \cite{FZ06}.
%%%%%%%%%%%%%%%%%%%%%%%%%%%%%%%%%%%%%%%%%%%%%%%%%%%%%%%%%%%%%%%%%%%%%%%%%%%%%%%%%%%%%%%%%%%%%%%%%%%%%%%%5
\section{Beyond the two-quark approximation \label{MQ}}
Eigenvalues $ \Delta \widetilde{E}_n(P) $ of the Bethe-Salpeter equation (\ref{BSP}) are not the 
same as the eigenvalues $ \Delta {E_n}(P) $ of the initial problem (\ref{eig}):
\begin{equation}\label{dEq}
\Delta {E_n}(P) = \Delta \widetilde{E}_n(P) + \delta {E_n}(P) .
\end{equation}
The difference $\delta {E_n}(P)$ is caused by the multi-quark corrections, which are ignored in (\ref{BSP}),
but contribute to $\Delta {E_n}(P)$. The exact meson energy spectra should have the form 
\begin{equation} \label{LorE}
\Delta E_n(P)=(M_n^2+P^2)^{1/2}.
\end{equation}
due to the Lorentz covariance of IFT, but this form does not hold \cite{FZ06} for the meson energies 
$\Delta \widetilde{E}_n(P)$ determined in the two-quark approximation. 

In the $P\to\infty$ limit equation (\ref{dEq}) yields due to (\ref{PInf}) and  (\ref{LorE}):
\[
M_n^2 =  \widetilde{M}_n^2 + \delta {M_n^2}, 
\]
where
\[
\delta M_n^2 = \lim_{ P\to \infty} [2 P\,\delta {E_n}(P)]. 
\]

The first analysis of the multi-quark corrections to the meson masses has been done by 
Fonseca and A.B.~Zamolodchikov \cite{FZ06}. They claim, that multi-quark corrections 
treated perturbatively in $\lambda$ should
modify the Bethe-Salpeter equation (\ref{BSP}) to the form
\begin{eqnarray} \label{BSPR}
\fl \left[ \varepsilon(P/2+p) +\varepsilon(P/2-p)- \Delta {E}(P)   \right] \Psi_P(p )     
=f\, \fint_{-\infty}^\infty 
 \mathbb{G}_P(p|k) \, \Psi_P(k) \, \frac{\rmd k}{2 \pi}.   
\end{eqnarray}
Here $\varepsilon(p)$ and $f$ are the renormalized quark dispersion law and the renormalized  string tension, respectively. 
The renormalized kernel $ \mathbb{G}_P(p|k)$ is assumed to have the structure
\begin{equation}\label{Rker}
 \mathbb{G}_P(p|k) = G_P(p|k) + \Delta \mathbb{G}_P^{(reg)}(p|k),
\end{equation}
where $G_P(p|k)  $ is the original kernel (\ref{G0}), and the correction term 
$ \Delta \mathbb{G}_P^{(reg)}(p|k) =O(\lambda)$,  being regular 
at $k=\pm p$, effectively modifies the pair interaction between quarks at short
distances $\lesssim m^{-1}$.

Note, that the renormalized quark energy does not have the Lorentz covariant form \cite{FZ06}, 
\begin{equation*}
\varepsilon(p) = (p^2+m^2)^{1/2}+ \delta \varepsilon(p) = (p^2+m_q^2)^{1/2}+ \Delta \varepsilon(p), 
\end{equation*}
 since quarks are not free particles at $h>0$ due to their confinement. Assuming $\Delta\varepsilon(p)=O(|p|^{-3})$
at  $p\to\infty$, one can define the "dressed" quark mass $m_q$ from the large-$p$ asymptotics of $\varepsilon(p)$:
\begin{equation*}
\varepsilon(p)=|p|+\frac{m_q^2}{2|p|}+ O(|p|^{-3}).
\end{equation*}
There are no nonperturbative definitions of renormalized quantities in equation (\ref{BSPR}). Instead, it is expected, that 
they can be determined order by order  by their power series in $\lambda$:
\numparts 
\begin{eqnarray} \label{mqs}
m_{q}^2 = m^2\,\big(1 + a_2\,\lambda^2 + a_3\,\lambda^3 + \cdots\big), \\ 
\delta \varepsilon(p) = \delta_2 \varepsilon(p)
+\delta_3 \varepsilon(p)   +O(\lambda^4),\label{dep}\\
f = f_0\,\big(1 + c_2\,\lambda^2 + c_4\,\lambda^4 + \cdots\big)\, \label{fex}\\
\Delta \mathbb{G}_P^{(reg)}(p|k) = 
\Delta_1 \mathbb{G}_P^{(reg)}(p|k)  + \Delta_2 \mathbb{G}_P^{(reg)}(p|k)+O(\lambda^3). \label{deG}
\end{eqnarray}
\endnumparts

Let us summarize briefly, what is known about the coefficients in the above expansions.
Fonseca an A.B.~Zamolodchikov \cite{FZWard03} analyzed the  exact integral 
representation for the coefficient $a_2$ in (\ref{mqs}),
and obtained from it the  value 
\begin{equation} \label{a2}
 a_2=0.071010809\ldots
\end{equation}
On the other hand, one can expand $a_2$ into the sum 
\begin{equation*}
a_{2} = a_{2,3} + a_{2,5} +\ldots
\end{equation*}
of the second order (in $\lambda$) diagrams with three, five, \ldots, quarks in the 
intermediate state. Contribution of three-quark diagrams into $a_2$ was estimated in
reference \cite{FonZam2003}
\begin{equation}  \label{a23FZ}
a_{2,3}\approx 0.07 \ldots
\end{equation}
We obtain  its exact value  
\begin{equation} \label{a23}
a_{2,3}=\frac{1}{16}+\frac{1}{12\pi^2}=0.07094\ldots,
\end{equation}
this calculation will be presented elsewhere.
Comparison of (\ref{a23}) with (\ref{a2}) shows, that the second order radiative correction to the 
quark mass is essentially
determined by the three-quark contribution. Diagrams with five and more quarks in the intermediate state
give less than 0.1 \% of $a_2$. 

The  term of order $\lambda^2$ in expansion (\ref{dep}) for $\delta \varepsilon(p)$ 
was found by Fonseca and A.B.~Zamolodchikov \cite{FZ06}:
\begin{equation} \label{d2eps}
 \delta_2\varepsilon(p)=\frac{\lambda^2 }{2}\frac{ m^2 a_2}{\omega(p)}-\frac{\lambda^2 }{8}\frac{m^4 p^2}{\omega^5(p)}.
\end{equation}
They  have given also strong arguments,
that coefficients $c_{2 k}$ in expansion (\ref{fex}) should be simply 
related with coefficients $\tilde{g}_{j}$ in the 
well known weak-$h$ expansion \cite{Berg79} for the vacuum energy $E_{vac}$, 
\begin{equation} \label{Evac}
E_{vac} = L\,m^2\left( 
-\,{1\over 2}\,\lambda + {\tilde g}_2\,\,\lambda^2 + {\tilde
  g}_3\,\,\lambda^3 + {\tilde g}_4\,\,\lambda^4 + \ldots\ \right),
\end{equation}
namely
\begin{equation}  \label{cg}
c_{2k} = -2\,{\tilde g}_{2k+1}.
\end{equation}
In particular, $c_2=-0.003889\ldots$

It is not difficult to modify the weak coupling expansions  (both low-energy and semiclassical) to 
account renormalized quantities in the  Bethe-Salpeter equation (\ref{BSPR}), and to the 
express multi-quark correction $\delta M_n$ in terms of coefficients in (\ref{mqs})-(\ref{deG}). 
It turns out \cite{FZ06}, that for calculation of the meson masses $M_n$ to the third order in $\lambda$, 
it would be  sufficient to know the renormalized quark mass $m_q$ and the string tension $f$ 
to the third order in $\lambda$, and the "regular" term  $\Delta \mathbb{G}_P^{(reg)}(p|q)$ 
in (\ref{Rker}) to the linear order in $\lambda$  in the limit $P\to\infty$. To this end,
one needs to determine two unknown quantities: the third order correction to the quark mass (coefficient $a_3$ in 
(\ref{mqs})), and the kernel $\Delta_1\mathbb{G}_\infty^{(reg)}(p|k)$ in (\ref{deG}). In fact, we need only the
diagonal part of the latter, $\Delta_1\mathbb{G}_\infty^{(reg)}(p|p)$. 

The problem of explicit calculation of $a_3$ and $\Delta_1\mathbb{G}_P^{(reg)}(p|k)$ is quite difficult. 
Here we do not try to find its complete solution. Instead, in subsequent sections we 
shall obtain several representations for these quantities in terms of formfactors of  
spin operators $\sigma(x)$ and their products $\sigma(x_1)\sigma(x_2)$. 
%%%%%%%%%%%%%%%%%%%%%%%%%%%%%%%%%%%%%%%%%%%%%%%%%%%%%%%%%%%%%%%%%%%%%%%%%%%%%%%%%%%%%%
\section{Diagonalization of the Hamiltonian in the fermionic number \label{DIAH}}
Bethe-Salpeter equation (\ref{BSP}) is approximate, since the IFT 
Hamiltonian (\ref{Ham}) does not conserve the number of fermions - the "bare" quarks. 
Let us try to find  a unitary operator $U(h)$, which transfers  operators generating
 "bare" fermions into operators generating such "dressed" fermions, that their number would be conserved
by the evolution operator. It is clear, that the two-fermion Bethe-Salpeter equation,
written for these "dressed" fermions, should be exact, and it could be identified with the 
renormalized Bethe-Salpeter equation (\ref{BSPR}).

Let $\underline{{\bf a}}^\dagger(p) ,\,\underline{{\bf a}}(p) $ be the set of 
creation/annihilation operators of the "dressed" fermions, which are
related with  the "bare" ones by the unitary transform  
\[
{\bf a}(p)=U(h) \,\underline{{\bf a}}(p) \,U(h)^{-1}, \quad  {{\bf a}}^\dagger(p)=U(h)\, \underline{{\bf a}}^\dagger(p) \,U(h)^{-1}
\]
with operator $U(h)$ depending on the magnetic field $h$. We shall also underline all 
"dressed"  operators and states: 
\begin{equation*}
\underline{A}=U(h)^{-1} \,A\,U(h) , \qquad  |\underline{\Phi}\rangle =U(h)^{-1} |{\Phi}\rangle.
\end{equation*}
Expanding $U(h)$ into the power series in $h$
\begin{equation*}
U(h)=1+\sum_{n=1}^\infty h^n\,\mathcal{F}_n, 
\end{equation*}
we obtain the set of equalities following from the unitarity condition $U(h)\,U(h)^\dagger =1$: 
\begin{eqnarray*}
&&\mathcal{F}_1+\mathcal{F}_1^\dagger=0,\\
&&\mathcal{F}_2=\frac{\mathcal{F}_1^2}{2}+\Lambda,\qquad \Lambda^\dagger=-\Lambda,\\
&&\mathcal{F}_3=\frac{\Lambda \mathcal{F}_1+\mathcal{F}_1 \Lambda}{2}+Y,\qquad Y^\dagger=-Y,\\
&& \quad\quad \ldots
\end{eqnarray*}
Denote by $\underline{N}$ the operator of the number of "dressed" fermions
\[
\underline{N}=\int_{-\infty}^\infty \frac{\rmd p}{2\pi}\,\underline{{\bf a}}^\dagger(p)\,\underline{{\bf a}}(p),
\]
and by $ \underline{\mathcal{P}}_{\,n} $ the projector operators onto the  subspaces
of $n$ "dressed" fermions.
For an operator $A$ acting in the Fock space let us separate the diagonal and off-diagonal parts 
in the "dressed" fermion number $n$,  $A=A_d+A_s$, where
\[
A_d = \sum_{n=0}^\infty\underline{\mathcal{P}}_{\,n} \,A \underline{\mathcal{P}}_{\,n},\qquad {\rm and}\qquad   A_s=A-A_d.
\]
We require, that $\mathcal{H}$ and $\underline{N}$ commute, $[\mathcal{H},\underline{N} ]=0$, 
or, equivalently, 
\begin{equation}
\mathcal{H}_s=0. \label{di}
\end{equation}
Rewriting (\ref{di}) as
\begin{equation*}
\mathcal{H}_s =\left(U(h)\underline{\mathcal{H}} U(h)^{-1}\right)_s =0,
\end{equation*}
one obtains
\begin{equation}
\fl \bigg(\big(1+h\mathcal{F}_1+ h^2\mathcal{F}_2+h^3\mathcal{F}_3+... \big)(\underline{\mathcal{H}}_0+h\underline{V})
\big(1+h\mathcal{F}_1^\dagger+ h^2\mathcal{F}_2^\dagger+h^3\mathcal{F}_3^\dagger+... \big)\bigg)_s=0. \label{expan}
\end{equation}

Let us collect linear in  $h$ terms in (\ref{expan}):
\begin{equation} \label{mcF}
\langle \underline{p}|\mathcal{F}_1|\underline{k}\rangle =
\frac{\langle \underline{p}|\underline{V}|\underline{k}\rangle}
{\omega(p)-\omega(k)} \quad \textrm{for} \quad  n(p)\ne n(k) .
\end{equation}
From here on we  use compact notations
$| \underline {k}\rangle =|\underline {k_1,...,k_{n(k)}}\rangle$, $\langle \underline {p}|=
\langle \underline {p_{n(p)},...,p_1}|$,
$\omega (p)= \omega (p_1)+...+\omega(p_{n(p)})$,  and so on. 
Equation (\ref{mcF}) defines $(\mathcal{F}_1)_s$, but does not impose restrictions on $(\mathcal{F}_1)_d$. 
We fix the latter by the condition $(\mathcal{F}_1)_d=0$.

In the second order in  $h$ one finds  from (\ref{expan}):
\begin{equation} \label{Lambda}
\bigg(\frac{\mathcal{F}_1^2}{2}\underline{\mathcal{H}}_0+\underline{\mathcal{H}}_0 \frac{\mathcal{F}_1^2}{2}
- \mathcal{F}_1  \underline{\mathcal{H}}_0 \mathcal{F}_1 +[\Lambda, \underline{\mathcal{H}}_0 ]+[\mathcal{F}_1,\underline{V}]
\bigg)_s=0.
\end{equation}
This equation defines $\Lambda_s$. We put $\Lambda_d=0$, and insert the intermediate state decomposition
\[
\fl 1=\sum_q  | \underline{q}\rangle\langle \underline{ q}|\equiv| \underline{0} \rangle\langle \underline{0}|+
\sum_{n(q)=1}^\infty\frac{1}{n(q)!}\int_{-\infty}^\infty 
|\underline{q}_{n(q)},\dots, \underline{q}_1 \rangle \langle \underline{q}_1,\ldots,\underline{q}_{n(q)}|
\,\prod_{j=1}^{n(q)} \frac{\rmd q_j}{2\pi}
\]
into (\ref{Lambda}), providing
\begin{eqnarray} 
\fl \langle \underline{p}|\Lambda|\underline{k}\rangle =
\frac{1}{\omega(k)-\omega(p)}\Bigg\{\sum_{q \atop {n(p)\ne n(q)\ne  n(k)}} 
\frac{\langle \underline{p}|\underline{V}|\underline{q}\rangle 
\langle \underline{q}|\underline{V}|\underline{k}\rangle}{[\omega(q)-\omega(p)][\omega(q)-\omega(k)]} 
 \left[\omega(q)- \frac{\omega(p)+\omega(k)}{2} \right] \nonumber \\+ 
 \sum_{q\atop{n(q)=n(k)}}\frac{\langle \underline{p}|\underline{V}|\underline{q}\rangle
 \langle \underline{q}|\underline{V}|\underline{k}\rangle}{[\omega(q)-\omega(p)]}+
\sum_{q\atop{n(q)=n(p)}}\frac{\langle \underline{p}|\underline{V}|\underline{q}\rangle 
\langle \underline { q}|\underline{V}|\underline{k}\rangle}{[\omega(q)-\omega(k)]}%\right]
\Bigg\} \label{La}\\
\fl {\rm for}\quad n(p)\ne n(k). \nonumber
\end{eqnarray}
Note, that one can drop all underlining in the right-hand sides of equations (\ref{mcF}) and (\ref{La}),  
since $\langle \underline {\Phi'}|\underline {A}|\underline {\Phi}\rangle =\langle \Phi'|A|\Phi\rangle$.
Similarly, we put $Y_d=0$, since equation (\ref{expan}) (in the third order in $h$) determines $Y_s$ only.

In the rest of this section we shall consider, how the Hamiltonian $\mathcal{H}$ acts in the 
subspaces with zero, one, and two renormalized fermions.
%%%%%%%%%%%%%%%%%%%%%%%%%%%%%%%%%%%%%%%%%%%%%%%%%%%%%%%%%%%%%%%%%%%%%%%%%%%%%%%%%%%%%%%%%%%%%%%%%%%%%%%%%%%%%%%%5
\subsection{Vacuum sector}
In the vacuum sector, one obtains from (\ref{mcF}), (\ref{Lambda}) the standard 
Rayleigh-Schr{\"o}dinger expansion  (\ref{Evac}) for the IFT ground state energy:
\begin{eqnarray} \nonumber
\fl E_{vac}= \langle  \underline{0}| \mathcal{H}| \underline{0}\rangle=
\langle  \underline{0}| U(h)( \underline{\mathcal{H}}_0+h \underline{V})U(h)^{-1}| \underline{0}\rangle=
h\langle  {0}| {V}| {0}\rangle + \delta_2 E_{vac} +\delta_3 E_{vac}++O(h^4),
\end{eqnarray}
where
\begin{eqnarray} \label{E2vac}
\fl \delta_2 E_{vac}=-h^2\sum_{q\atop{n(q)\ne 0}}
\frac{\langle  {0} | {V}|q\rangle \langle q|  {V}| {0}\rangle}
{\omega(q)}, \\
\fl\delta_3 E_{vac}=+h^3\Bigg\{-\langle  {0}| {V}| {0}\rangle \sum_{q\atop{n(q)\ne 0}}
 \frac{\langle  {0}| {V}|q\rangle \langle q| {V}| {0}\rangle}
{[\omega(q)]^2} 
 +\sum_{q,q'\atop{n(q)\ne 0\ne n(q')}}\frac{\langle  {0}| {V}|q\rangle\,
\langle q| {V}| {q'}\rangle
\, \langle q'| {V}| {0}\rangle}{\omega(q)\omega(q')}
\Bigg\}. \nonumber
\end{eqnarray}
%%%%%%%%%%%%%%%%%%%%%%%%%%%%%%%%%%%%%%%%%%%%%%%%%%%%%%%%%%%%%%%%%%%%%%%%%%%%%%%%55
\subsection{One-fermion sector}
In the one-fermion sector $n(p)=n(k)=1$, and we find
\begin{equation} \label{1sect}
 \fl \langle \underline{p}|\mathcal{H}|\underline{k}\rangle = 2\pi \delta(p-k)\,\omega(p)+h \langle p| {V}|k\rangle
+ \delta_2 \langle \underline{p}|\mathcal{H}|\underline{k}\rangle +\delta_3 \langle \underline{p}|\mathcal{H}|\underline{k}\rangle
+O(h^4),
\end{equation}
where
\begin{eqnarray} \label{delta2}
\fl \delta_2 \langle \underline{p}|\mathcal{H}|\underline{k}\rangle=  -
\frac{h^2}{2}\,\sum_{q\atop {n(q)\ne n(p)}}\langle p | {V}|q\rangle \langle q|  {V}|k\rangle 
\left[\frac{ 1 }{\omega(q)- \omega(p)} +\frac{ 1 }{\omega(q)- \omega(k)}\right], \\
 \fl \delta_3 \langle \underline{p}|\mathcal{H}|\underline{k}\rangle=    \label{delta3}
+\frac{h^3}{2}\sum_{q,q'}  
\langle p | {V}|q\rangle \langle q|  {V}|q'\rangle\langle q'|  {V}|k\rangle \Bigg\{ 
[1-\delta_{n(q),n(p)}][1-\delta_{n(q'),n(p)}]\\
\fl \cdot\Bigg[\frac{1}{[\omega(p)-\omega(q)]}\frac{1}{[\omega(p)-\omega(q')]}  
 +\frac{1}{[\omega(k)-\omega(q)]}\frac{1}{[\omega(k)-\omega(q')]}\Bigg] \nonumber \\ 
\fl + \frac{1}{\omega(q)-\omega(q')}\left[\frac{\delta_{n(q),n(p)}[1-\delta_{n(q'),n(p)}]}{\omega(q')-\omega(p)} - 
\frac{[1-\delta_{n(q),n(p)}]\delta_{n(q'),n(p)}}{\omega(q)-\omega(k)}\right]\Bigg\}   \nonumber
\end{eqnarray}
First, let us consider the linear term in $h$ in the right-hand side of (\ref{1sect})
\begin{equation}\label{V1}
h \langle p| {V}|k\rangle =-h\int_{-\infty}^\infty \rmd {\rm x}\,  \langle p|\sigma({\rm x})|k\rangle,
\end{equation}
where
\begin{equation*}
\langle p | \sigma({\rm x}) |k\rangle = \frac{ \rmi\,\bar{\sigma}\, \exp[\rmi  {\rm x}(k-p)]   }{p-k}\,
\frac{ \omega(p)+\omega(k) }{[\omega(p)\omega(k)]^{1/2}}
\end{equation*}
is the formfactor of the order spin operator (\ref{sigma}) in the momentum basis. 
Integration in ${\rm x}$ in (\ref{V1}) leads to the divergent result 
\begin{equation} \label{Vdiv}
h \langle p| {V}|k\rangle=-2\pi \rmi \delta(p-k) \frac{f_0}{p-k}.
\end{equation}
This  singularity is well known in the standard formfactor perturbative theory, 
where it  appears as the divergency of the first order correction to the fermion mass,
which is interpreted as a formal indication of confinement  \cite{Del96,Del04}. 
  
To give a meaning to equation (\ref{Vdiv}), let us  mention, that the 
generalized function $\delta(q)/q$ is well defined 
and equivalent to $-\delta'(q)$ in the class of the main functions 
$\varphi(q)\in C^1$
taking zero value at the origin, $\varphi(0)=0.$ So, one can formally write
\[
\delta(p-k) \frac{f_0}{p-k}=-\delta'(p-k)+C\delta(p-k)
\]
with some indeterminate constant $C$. 

To get further insight, it is instructive to
consider  the matrix element $ \langle {\rm X}|h\, {V}|k\rangle$, where the state
$ \langle {\rm X}|$ describes a "bare" quark located at the point ${\rm X}$:
\begin{equation} \label{kinkl}
\langle {\rm X}|=\int_{-\infty}^\infty\frac{\rmd p}{2\pi}\e^{\rmi p {\rm X}} \langle p|.
\end{equation} 
\begin{figure}[ht]
\centering
\includegraphics[width= .8\linewidth]{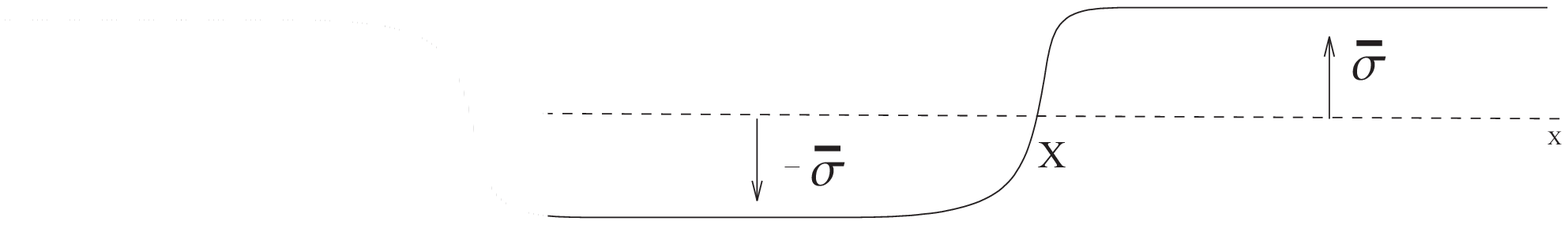}
\caption{One-fermion state (\ref{kinkl}) represents a kink centred at ${\rm X}$. \label{kink}} 
\end{figure}
For the matrix element of the order spin operator $\sigma({\rm x}) $ we  get
\begin{eqnarray}  \label{formp}
\fl \langle {\rm X} | \sigma({\rm x}) |k\rangle = 
  \fint_{-\infty}^\infty  \frac{\rmd p}{2 \pi }\, \rmi\,\bar{\sigma }
 \frac{\exp[\rmi p ({\rm X}-{\rm x})+\rmi k {\rm x}] }{p-k}\,
\frac{ \omega(p)+\omega(k) }{\sqrt{\omega(p)\omega(k)}} \nonumber\\ 
\fl 
= {\rm sign} ({\rm x}-{\rm X}) \,\bar{\sigma }\, \rme^{\rmi k {\rm X}}+\rmi \,\bar{\sigma }\rme^{\rmi k {\rm x}}
\int_{-\infty}^\infty \frac{\rmd p}{2 \pi}\frac{\exp[\rmi p ({\rm X}-{\rm x})] }{p-k}\,\
\left\{ \left[ \frac{ \omega(p) }{\omega(k)} \right]^{1/4} - 
\left[ \frac{ \omega(k) }{\omega(p)} \right]^{1/4}\right\}^{2}, \label{sig}
\end{eqnarray}
where $\fint$ denotes the Cauchy principal value integral. Here the first term in the right-hand side  is 
non-local, while the second term  is well localized 
near the diagonal ${\rm x}={\rm X}$  exponentially vanishing for $| {\rm x} -{\rm X}|\gg m^{-1}$. 
Equation (\ref{sig}) allows one to interpret the one-fermionic state $\langle {\rm X}|$ 
as a kink of width $\sim m^{-1}$ centred at ${\rm X}$, which devides  the regions 
with magnetizations  $-\bar{\sigma}$ to the
left, and  $+\bar{\sigma}$ to the right side of it, see figure~\ref{kink}.

Substitution of (\ref{sig}) into (\ref{V1}) yields after integration in ${\rm x}$
\begin{equation*}
h \langle {\rm X} |V|k\rangle = (f_0 \,{\rm X}  +  C  )\,\rme^{\rmi k {\rm X}},
\end{equation*}
where $f_0=2 h \bar{\sigma } $ is the "bare string tension" , 
and the constant $C$ is proportional to the length of the system
$L$ being infinite  in the thermodynamic limit. Thus, Hamiltonian (\ref{Ham}) acts in the one-particle 
subspace  of "bare" quarks ${\mathbb F}_1$ as
\begin{equation*} %\label{Ham1}
{\mathcal P}_1 {\mathcal H} {\mathcal P}_1  = \omega(\hat{p})+ f_0\,\hat{ {\rm x} } + C, 
\end{equation*}
where $\hat{ {\rm x} }$ and $ \hat{  p } $ are the one-particle coordinate and momentum operators. 
The same formula written for "dressed" quarks
\begin{equation} \label{Hamd}
\underline{\mathcal P}_1 {\mathcal H} \underline{\mathcal P}_1  = 
\varepsilon(\hat{p})+ f\,\hat{ {\rm x} } + {C}_R 
\end{equation}
gives us the  perturbative definition of the renormalized quark dispersion law $\varepsilon(p)$
and renormalized string tension $f$. In the momentum representation (\ref{Hamd}) takes the form
\begin{equation*}
 \langle \underline{p}|\mathcal{H}|\underline{k}\rangle =
 2\pi \delta(p-k)\,[\varepsilon(p)+C_R]+2\pi \rmi \,f\,\delta'(p-k), 
\end{equation*}
which should be compared with (\ref{1sect}) order by order in $h$.

The second order term in (\ref{1sect}) determines the leading correction to  the quark energy $\delta_2 \varepsilon(p)$
in expansion (\ref{dep})
\[
\delta_2 \langle \underline{p}|\mathcal{H}|\underline{k}\rangle =
 2\pi \delta(p-k)\,[ \delta_2\varepsilon(p)+ \delta_2 E_{vac} ], 
\]
where $ \delta_2 E_{vac} $ is given by (\ref{E2vac}).  Explicitly, it can be described either by the 
formfactor expansion following from (\ref{delta2})
\begin{eqnarray}\label{e2form}
 \delta_2\varepsilon(p) = \delta_{2,3}\varepsilon(p) +\delta_{2,5}\varepsilon(p)+\ldots, \\
 \delta_{2,n}\varepsilon(p) = -\frac{h^2}{n!}
\int_{-\infty}^\infty\frac{\rmd q_1\ldots \rmd q_{n}}{(2\pi)^{n-1}}\frac{\delta(q_1+\ldots+q_{n}-p)} 
{\omega(q_1)+\ldots+\omega(q_{n})-\omega(p)}  \label{epsn}\\
\cdot \lim_{k\to p} \langle p|\sigma(0)|q_1,\ldots,q_{n}\rangle \langle q_{n},\ldots, q_1|\sigma(0)|k\rangle,  \nonumber
\end{eqnarray}
or by the equivalent integral representation 
\begin{equation} \label{mint}
 \delta_2\varepsilon(p) =-h^2\int_{-\infty}^\infty \rmd {\rm x} \int_{0}^\infty \rmd {\rm y}
\,\lim_{k\to p}\,\langle p | \sigma({\rm x, y}) (1-\mathcal{P}_1)\sigma(0,0)|k\rangle,
\end{equation}
where $\sigma({\rm x, y})= \exp(-\rmi{\rm x} \hat{P}  +{\rm y} \mathcal{H}_0) 
\sigma(0)\exp(\rmi{\rm x} \hat{P} -{\rm y} \mathcal{H}_0)$.

Representations (\ref{e2form})-(\ref{mint}) were first obtained and studied by 
Fonseca and A.B.~Zamolodchikov \cite{FonZam2003,FZWard03, FZ06}, we quoted their results in section \ref{MQ}
[see equations (\ref{a2}), (\ref{a23FZ}), (\ref{d2eps})]. We  determine the exact large-$p$ 
asymptotics of the integral (\ref{epsn})  for $n=3$
\begin{equation}
 \delta_{2,3}\varepsilon(p)=\frac{\lambda^2 m^2}{2 p} \left(\frac{1}{16}+\frac{1}{12\pi^2}\right)+O(p^{-2}),
\end{equation}
which leads to (\ref{a23}).

The third order term (\ref{delta3}) in (\ref{1sect}) contributes 
both to the string tension $f$, and to the quark energy $\varepsilon(p)$.
It determines $\delta_3\varepsilon(p)$ and the constant  $a_3$ in  expansion (\ref{mqs}) for the quark mass $m_R$. 
Calculation of $a_3$ would be of much interest for interpreting recent numerical calculation 
of masses of lightest mesons,  see figure~7 in \cite{FZ06}. 
%%%%%%%%%%%%%%%%%%%%%%%%%%%%%%%%%%%%%%%%%%%%%%%%%%%%%%%%%%%%%%%%%%%%%%%%%%%%%%%%%%%%%%%%%%%%%%%%%%%%%
\subsection{Two-fermion sector}
In the two-particle sector of "dressed" quarks, Hamiltonian acts as
\begin{eqnarray} \nonumber
\fl \langle \underline{ p_2,p_1}|\mathcal{H}|\underline{k_1,k_2}\rangle = (2\pi)^2[\omega(p_1)+\omega(p_2)][
 \delta(p_1-k_1)\delta(p_2-k_2)- \delta(p_1-k_2)\delta(p_2-k_1)]\\
 +h \langle p|V|k\rangle + \delta_2 \langle \underline{p}|\mathcal{H}|\underline{k}\rangle +
\delta_3 \langle \underline{p}|\mathcal{H}|\underline{k}\rangle,
+O(h^4)\label{2sect} 
\end{eqnarray}
where $\delta_2 \langle \underline{p}|\mathcal{H}|\underline{k}\rangle$ 
and $\delta_3 \langle \underline{p}|\mathcal{H}|\underline{k}\rangle$ 
are given by equations (\ref{delta2}), (\ref{delta3}) with $n(p)=n(k)=2$. 
Two initial terms in the right-hand side of (\ref{2sect}) give rise to the "bare" Bethe-Salpeter equation
(\ref{BSP}).

The explicit form of the second-order correction is 
\begin{equation*} 
 \delta_2 \langle \underline{{p}_2 ,{p}_1} | \mathcal{H}|\underline{k_1 ,k_2}\rangle=
-4 \pi \,f_0 \,\delta(p_1+p_2-k_1-k_2)\,\delta_2 \mathcal{G}(p_1,p_2|k_1,k_2),
\end{equation*}
where
\begin{eqnarray} \label{H2}
\fl \delta_2 \mathcal{G}(p_1,p_2|k_1,k_2) =\frac{h}{8\bar{\sigma}}\sum_{j=2}^{\infty} \frac{1}{(2 j)!}\int_{-\infty}^\infty
\frac{\rmd q_1\ldots \rmd q_{2j}}{(2\pi)^{2j-1}}\delta(p_1+p_2-q_1-\ldots-q_{2j})\\
\fl\cdot\langle p_2, p_1|\sigma(0)|q_1,\ldots,q_{2 j}\rangle \langle q_{2 j},\ldots, q_1|\sigma(0)|k_1, k_2\rangle  \nonumber\\
\fl\cdot\left[ \frac{1}{\omega(q_1)+\ldots+\omega(q_{2j})- \omega(p_1) -\omega(p_2)} +
\frac{1}{\omega(q_1)+\ldots+\omega(q_{2j})- \omega(k_1) -\omega(k_2)}\right].  \nonumber
\end{eqnarray}

Application of the Wick expansion to the formfactors in the integrand brakes (\ref{H2}) into the sum of diagrams. 
Some of them contain one or two products of the form
\begin{equation}
\fl\langle p_\alpha|\sigma(0)|q_\gamma\rangle  \langle q_\gamma|\sigma(0)|k_\beta\rangle= \frac{\omega(p_\alpha)+\omega(q_\gamma)}
{[\omega(p_\alpha)\omega(q_\gamma)]^{1/2}}\frac{\omega(k_\beta)+\omega(q_\gamma)}
{[\omega(k_\beta)\omega(q_\gamma)]^{1/2}}\,
\mathcal{P}\frac{1}{p_\alpha-q_\beta}\cdot\mathcal{P}\frac{1}{k_\beta-q_\beta}, \label{pv}
\end{equation} 
which have two kinematic singularities in the integration variable $q_\gamma$ 
at $q_\gamma=p_\alpha$, and $q_\gamma=k_\beta$. 
Here $\mathcal{P}\frac{1}{z}$ denotes the "principal value" generalized function,
\[
\mathcal{P}\frac{1}{z} =\frac{1}{2}\left( \frac{1}{z+\rmi 0} +\frac{1}{z-\rmi 0}\right).
\]
Let us rewrite the singular factor in the right-hand side of (\ref{pv}) as
\begin{equation} \label{pv2}
\fl \mathcal{P} \frac{1}{p_\alpha-q_\gamma} \cdot\mathcal{P} \frac{1}{k_\beta-q_\gamma} =
\mathcal{P}\frac{1}{(p_\alpha-q_\gamma)(k_\beta-q_\gamma)}
+\pi^2 \delta(p_\alpha-k_\beta)\delta(p_\alpha-q_\gamma),
\end{equation}
where
\begin{equation*}
\fl \mathcal{P}\frac{1}{(p_\alpha-q_\gamma)(k_\beta-q_\gamma)} =
\frac{1}{2}\left[  \frac{1}{(p_\alpha-q_\gamma-\rmi 0)(k_\beta-q_\gamma-\rmi 0) } +
\frac{1}{(p_\alpha-q_\gamma+\rmi 0)(k_\beta-q_\gamma+\rmi 0) } \right].
\end{equation*} 
Substitution of (\ref{pv2}) into factors (\ref{pv}) leads to splitting of diagrams containing
(one or two) such factors into  several (two or four) terms. The resulting diagrams can be classified by the number
of $\delta$-functions $\delta(p_\alpha-k_\beta)$ arising from the second term in the right-hand side of (\ref{pv2}). 
\begin{enumerate}
\item Diagrams with two
$\delta$-functions give rise to the vacuum energy correction $\delta_2 E_{vac}$. 
\item Diagrams with
one $\delta$-function contribute to the corrections $\delta_2 \varepsilon(p_1) $ , 
$\delta_2 \varepsilon(p_2)$ to the energies of two quarks. 
\item The rest diagrams are regular at $p_\alpha=k_\beta$ for  $\alpha, \beta=1,2$. We denote 
their sum by  $\delta_2 \mathcal{G}^{(reg)}(p_1,p_2|k_1,k_2) $. It determines 
(to the linear order in $h$) the kernel $\Delta \mathbb{G}_P^{(reg)}(p|q)$ in the renormalized 
Bethe-Salpeter equation (\ref{BSPR}): 
\begin{equation*} 
\fl\Delta_1 \mathbb{G}_P^{(reg)}(p|k) = \delta_2 \mathcal{G}^{(reg)}(P/2+p,P/2-p|P/2+k,P/2-k).
\end{equation*}
\end{enumerate}
%%%%%%%%%%%%%%%%%%%%%%%%%%%%%%%%%%%%%%%%%%%%%%%
\section{Local multi-quark corrections to the meson masses \label{MCC}}
It is not difficult to account perturbatively the regular correction term $\Delta_1 \mathbb{G}_P^{(reg)}(p|k)$ 
in the Bethe-Salpeter equation (\ref{BSPR}) both in the low-energy and  semiclassical expansions. 
The resulting local multi-quark correction to the meson energy reads as 
\begin{equation} \label{E3L}
\delta_3 E_n(P)=-\frac{f_0^2\,\omega^3(P/2)}{4m^2}\,\frac{\partial^2}{\partial p^2}\Bigg|_{p=0}
\Delta_1 \mathbb{G}_P^{(reg)}(p|p)
\end{equation}
in the low-energy case $n\sim 1$, and 
\begin{equation} \label{E3S}
\delta_3 E_n(P)=-\frac{2 \,f_0^2}{|(p_1-p_2)\,v|}\,\Delta_1 \mathbb{G}_P^{(reg)}
\left( \frac{p_1-p_2}{2}\bigg| \frac{p_1-p_2}{2}\right)
\end{equation}
in the semiclassical case $n\gg 1$. Here momenta $p_1$ and $p_2$ are the solutions of equations
\begin{eqnarray*}
p_1+p_2=P,\\
\omega(p_1)+\omega(p_2)=(P^2+M_n^2)^{1/2},
\end{eqnarray*}
and 
\[
v= \frac{p_1}{\omega(p_1)} -\frac{p_2}{\omega(p_2)}.
\] 

To obtain the local multi-quark corrections to the meson masses, we rewrite  (\ref{E3L}), (\ref{E3S}) in 
the rapidity variables $\beta_1={\rm arcsinh}(p_1/m)$, $\beta_2={\rm arcsinh}(p_2/m)$, and then proceed to the limit $P\to\infty$,
yielding   
\begin{equation} \label{M3low}
\frac{\delta_3 M_n^2}{m^2} =-\frac{\lambda^3}{8}\lim_{\beta_1\to\infty}\left( \frac{\partial}{\partial \beta_1} -
\frac{\partial}{\partial \beta_2} \right)^2\bigg|_{\beta_1=\beta_2} \frac{m^2 \,W(\beta_1,\beta_2)}{\bar{\sigma}^2}
\end{equation}
in the low-energy case, and 
\begin{equation}  \label{M3c1}
\frac{\delta_3 M_n^2}{m^2} =-\frac{\lambda^3\,m^2}{M_n^2-4 m^2}\lim_{\beta\to\infty} 
\,\frac{m^2 \,W(\beta+\eta,\beta-\eta)}{\bar{\sigma}^2}
\end{equation}
in the semiclassical case.
Here $\eta={\rm arccosh} [M_n/(2 m)]$, and 
\begin{eqnarray} \label{defW}
 W(\beta_1,\beta_2) = \frac{4 \bar{\sigma}}{h} \omega(p_1)\omega(p_2)
\,\Delta_1 \mathbb{G}_P^{(reg)}(p_1,p_2), 
\end{eqnarray}
where $p_j=m \sinh \beta_j$ for $j=1,2$. Function $ W(\beta_1,\beta_2)$ determined by (\ref{defW})
admits a compact integral representation, analogous to (\ref{mint}):
\begin{equation} 
\fl W(\beta_1,\beta_2)=  \int_{\infty}^\infty \rmd {\rm x}\int_{0}^\infty \rmd {\rm y}
\lim_{ \beta_{1}'\to\beta_{1} \atop \beta_{2}'\to\beta_{2}}
\langle\beta_2,\beta_1|\sigma({\rm x},{\rm y } )(1- \mathcal{P}_0-\mathcal{P}_2)\sigma(0,0)|\beta_1',\beta_2'\rangle.
\label{WW}
\end{equation}
In \ref{integ} we extract from this function the leading  at $(\beta_1+\beta_2)\to\infty$ "irreducible" part 
$W_{irr}(\beta_1,\beta_2)$  which determines $\delta_3 M_n^2$. It is expressed there in terms of the
two-fermion matrix elements of the  order spin operator pairs, which are explicitly known \cite{FZWard03}.

The third-order term $\delta_3 \langle \underline{p}|\mathcal{H}|\underline{k}\rangle$ in 
(\ref{2sect}) also contains regular and singular parts. The former contributes to the meson masses
only in the forth order. The latter renormalizes the quark dispersion law and the string tension, 
which give rise to the "nonlocal" third-order multi-quark correction 
to the meson energy $E_n(P)$. It is expected \cite{FZ06}, however, 
that {\it in the limit} $P\to\infty$ the nonlocal multi-quark corrections to $E_n(P)$ can be absorbed by renormalization 
of the meson mass and string tension.  Thus, the third-order multi-quark correction to the meson masses 
should be described by representations (\ref{M3low}) and (\ref{M3c1}), in which the "bare" parameters should be 
replaced by the renormalized ones $m\to m_q$, $\lambda\to\lambda_R$, where $\lambda_R=f/m_q^2$, and 
$m_q$ and $f$ are given by expansions (\ref{mqs}), (\ref{fex}).
%%%%%%%%%%%%%%%%%%%%%%%%%%%%%%%%%%%%%%%%%%%%%%%%%%%%%%%%%%%%%%%%%%%%%%%%%%%%%%%%%%%%%%%%%%
\section{Conclusions \label{CONC}}
This paper is devoted to extension to the third order of the weak-$h$ expansions for the meson  masses $M_n(h)$ in
the ferromagnetic IFT. There are four third-order contributions to it. 
The first one comes from the Bethe-Salpeter equation (\ref{BSi}) written in the two-quark approximation. 
For the semiclassical expansion this contribution is described by equations (\ref{2M})-(\ref{S2}), 
for the low-energy expansion
was already determined in reference \cite{FZ06}, see (\ref{lowseries}).
Three other contributions to $M_n(h)$ are due to the multi-fermion fluctuations. 
The local contribution arises from the regular radiative correction to the Bethe-Salpeter kernel. 
For this contribution, we obtain the integral representations (\ref{M3low})-(\ref{WW}), which are compact and 
appropriate for analytical analysis, and representation (\ref{M3}), (\ref{Wirr}), (\ref{Gir}), which  we plan to use 
in the future numerical calculations. The two last multi-quark contributions to 
the meson mass come from the third order corrections $\delta_3 \varepsilon(p)$ and $\delta_3 f$ to the quark self energy 
and  string tension, which are contained implicitly in the  formfactor expansion (\ref{delta3}). 
Explicit extraction of $\delta_3 \varepsilon(p)$ and $\delta_3 f$ from (\ref{delta3}) is rather involved. 
Whereas for $\delta_3 f$ the result is essentially known [see (\ref{fex}), (\ref{Evac}), 
(\ref{cg})], explicit calculation of the third order
correction to the quark self energy and quark mass remains an open problem. 
%%%%%%%%%%%%%%%%%%%%%%%%%%%%%%%%%%%%%%%%%%%%%%%%%%%%%%%%%%%%%%%%%%%%%%%%%%%%%%%%%
%%%%%%%%%%%%%%%%%%%%%%%%%%%%%%%%%%%%%%%%%%%%%%%%%%%%%%%%%%%%%%%%%%%%%%%%%%%%%%%%%%%%%%
\noindent
\ack 
This work 
is supported by  the Belarusian 
Republican Foundation for Fundamental Research under the grant ${\rm \Phi}$07-147. 
\appendix
\section{Perturbative solution of the Bethe-Salpeter equation \label{ExBS} }
In this Appendix we describe perturbative solution of the "bare" Bethe-Salpeter equation (\ref{BSi})
in the infinite momentum frame $P\to \infty$ to the third order in the magnetic field $h$. 
\subsection{Some exact relations}
It is convenient to rewrite equation (\ref{BSi}) in new notations
\begin{eqnarray*}
\phi(u)=\Phi(u) (1-u^2)^{-1/2}, \quad
\nu^2=\frac{ \widetilde M^2-4 m^2}{\widetilde M^2}, \quad
\rho =\frac{2 h \bar{\sigma}}{\widetilde M^2}.
\end{eqnarray*}
Since $\phi(-u)=-\phi(u)$, equation  (\ref{BSi}) takes the form:
\begin{equation}
(u^2-\nu^2) \phi(u) =\rho\fint_{-1}^1 \frac{\rmd v}{\pi}
\phi(v)\bigg[\frac{u v}{2}+4\frac{1-uv}{(u-v)^2}\bigg],\label{BS1}
\end{equation}
or equivalently
\begin{equation}
(u^2-\nu^2)\phi(u)=2 \rmi a \rho u+4 \rho \,[-u+(1-u^2)\,\partial_u] \fint_{-1}^1 \frac{\rmd v}{\pi}
\frac{\phi(v)}{v-u}, \label{BS2}
\end{equation}
where
\[
a=\int_{-1}^1 \frac{\rmd v}{4\pi \rmi}\,v\,\phi(v). 
\]
We shall require, that $\phi(u)$ is a purely imaginary function in the interval $(-1,1)$. 

Set 
\begin{equation}
g(u)=\frac{1}{2 \pi \rmi} \int_{-1}^1 \frac{\rmd v}{v-u}\,\phi(v), \label{g}
\end{equation}
for complex $u\notin[-1,1]$.
\begin{figure}[htbp]
  \centering
\includegraphics[width=\textwidth]{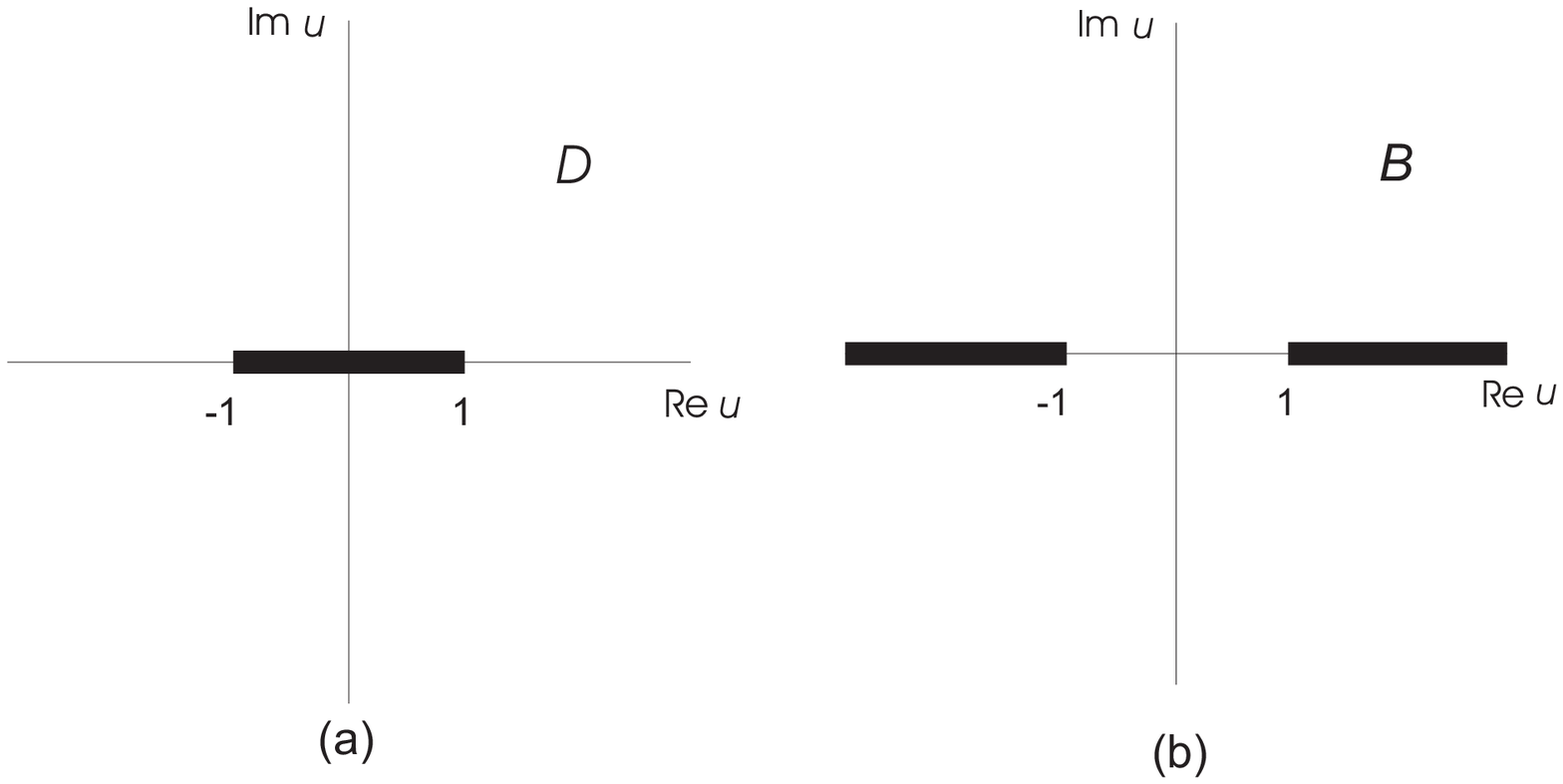}
  \caption{ (a) $D$ is the region of analyticity of  $g(u)$, 
     (b) $B$ is the region of analyticity of  ${\bf U}(u)$,  }
  \label{figD}
\end{figure}
Function $g(u)$ is analytical in the region $D$ shown in figure~\ref{figD}(a), and
\[
g(u)=-\frac{2 a}{u^2}+O(u^{-3}) \quad {\rm at}\; u \to \infty, \nonumber
\]
providing
\begin{equation}
2a={\rm Res}\mid _{u=\infty} [g(u)\,u]. \label{a}
\end{equation}
For real $u\in(-1,1)$ we get
\begin{eqnarray*}
\phi(u)=g(u+\rmi 0)-g(u-\rmi 0),\\
\fint_{-1}^1 \frac{\rmd v}{\pi }\,\frac{\phi(v)}{v-u} =\rmi [g(u+\rmi 0)+ g(u-\rmi 0)],
\end{eqnarray*}
and equation (\ref{BS2}) takes the form:
\begin{eqnarray} 
(u^2-\nu^2)[g(u+\rmi 0)-g(u-\rmi 0)] \label{gg}\\
=4 \rmi \rho \,[-u+(1-u^2)\,\partial_u] 
[g(u+\rmi 0)+ g(u-\rmi 0)]+2 \rmi a \rho u .  \nonumber
\end{eqnarray}
Let us define two function in $D$:
\begin{eqnarray}
\fl {\bf{U}}_1(u) &=&-4 \rmi \rho  [-u+(1-u^2)\,\partial_u]g(u)+(u^2-\nu^2)g(u)-\rmi a\rho u,
\label{eqU+} \\
\fl {\bf{U}}_2(u)&=&4 i \rho  [-u+(1-u^2)\,\partial_u]g(u)+(u^2-\nu^2)g(u)+\rmi a\rho u.  \nonumber
\end{eqnarray}
Due to (\ref{gg}), we have ${\bf{U}}_1(u+\rmi 0)={\bf{U}}_2(u-\rmi 0)$ for $u\in(-1,1)$. Therefore, function 
$ {\bf{U}}(u) $ defined as 
\[
{\bf{U}}(u) =\cases{ {\bf{U}}_1(u)\quad {\textrm {for Im}}\,u>0 , \\{\bf{U}}_2(u)\quad {\textrm {for Im}}\,u<0 }
\]
can be analytically continued into the 
 complex region $B$ shown in figure~\ref{figD}(b). Note, that ${\bf{U}}(u)$ is even in $B$, and real in the interval 
$(-1,1)$.

\begin{figure}[htbp]
  \centering
\includegraphics[width=.5\textwidth]{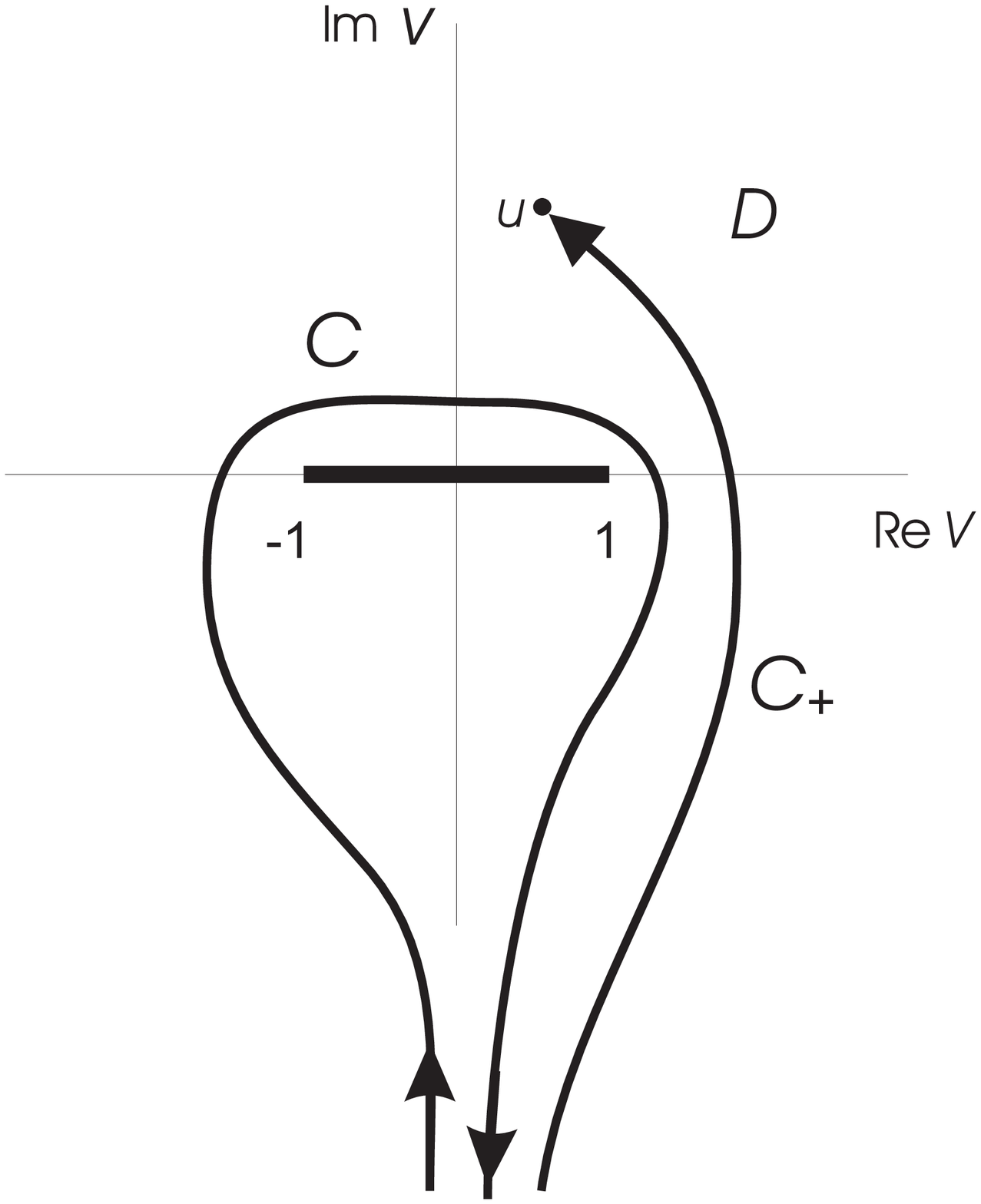}
  \caption{Integration paths in equations (\ref{pathC}), (\ref{req}).}
  \label{figC}
\end{figure}

Let us solve differential equation (\ref{eqU+}) with respect to $g(u)$:
\begin{equation}
\fl g(u)=\frac{\rmi}{4\rho}\,\int_{-\rmi\infty}^u \rmd v\,
\frac{{\bf{U}}_+(v)\exp\left[\frac{\rmi}{4\rho}(u-v)\right]}{[(1-u^2)(1-v^2)]^{1/2}}
\left[\frac{(1-u)(1+v)}{(1+u)(1-v)}\right]^{\rmi(1-\nu^2)/(8\rho)} ,\label{pathC}
\end{equation}
where ${\bf{U}}_+(u)={\bf{U}}_1(u)+\rmi a\rho u$, and  the branch of the last factor in the integrand is fixed as
\[
\arg\left[\frac{1-(u+\rmi 0)}{1+(u+\rmi 0)}\right]=\arg\left[\frac{1+(v+\rmi 0) }{ (1-(v+\rmi 0) }\right]=0\quad 
{\rm for\; real }\; u,v\in (-1,1). 
\]
Integration in (\ref{pathC}) goes along the path $C_+$  shown in figure~\ref{figC}. 

Function $g(u)$ should be single-valued in $D$. The trivial monodromy behaviour of $g(u)$ at $u=\infty$ 
is provided by (\ref{pathC}), if 
the following requirement is satisfied:
\begin{equation}
\int_C \rmd v\,\frac{{\bf{U}}_+(v)}{(1-v^2)^{1/2}}
\left[\frac{(1+v)}{(1-v)}\right]^{\rmi(1-\nu^2)/(8\rho)}\exp\bigg[-\frac{\rmi\,v}{4\rho}\bigg]
=0,\label{req}
\end{equation}
where the integration path $C$ is shown in figure~\ref{figC}.
The last condition determines the spectrum $\nu_n$. If $\rho\ll1$, the integral in (\ref{req})
is determined by contributions of the saddle points $v=\pm \nu$ of the function ${ \Upsilon }(v)$:
\begin{equation}
\Upsilon (v)=\frac{1-\nu^2}{2}\ln\bigg(\frac{1+v}{1-v}\bigg)-v.
\end{equation}
In the semiclassical limit $n\gg1$, contributions of these two saddle points are well separated, and (\ref{req})
yields the final asymptotical formula
\begin{equation}
\fl -\frac{ \Upsilon (\nu_n)}{4}=\rho \pi \left(n-\frac{1}{4}\right)+\rho \,\arg \left\langle
\frac{{\bf{U}}_+(\nu_n+\Delta v)}{\sqrt{1-(\nu_n+\Delta v)^2}}\exp\left[\frac{\rmi}{4\rho} 
\Delta\Upsilon(\nu_n+\Delta v)\right]\right\rangle,  \label{reqs}
\end{equation}
valid to all orders in $\rho\to 0$.
Here 
\[
\Delta\Upsilon (\nu+\Delta v) =\Upsilon(\nu+\Delta v)-\Upsilon(\nu)-\frac{\nu (\Delta v)^2}{1-\nu^2},
\]
averaging $ \langle \ldots\rangle $ stands for
\[
\fl \langle f(\Delta v)\rangle= \int_{-\infty}^\infty \rmd\,\Delta v \, f(\Delta v) 
\exp\left[\frac{\rmi\nu (\Delta v)^2}{4\rho(1-\nu^2)}  \right] \left\{
 \int_{-\infty}^\infty \rmd\,\Delta v' \, 
\exp\left[\frac{\rmi\nu (\Delta v')^2}{4\rho(1-\nu^2)}\right] 
\right\}^{-1},
\]
providing 
\begin{equation}
\fl \langle(\Delta v)^{2j+1} \rangle=0, \quad\quad 
 \langle(\Delta v)^{2j} \rangle=[4\rmi\rho(1-\nu^2)/\nu]^j \,
\Gamma(1/2+j)/\Gamma(1/2). \label{av}
\end{equation} 
At small $\rho, \Delta v$, function ${\bf{U}}_+(\nu+\Delta v)$ can be expanded as
\begin{equation}
{\bf{U}}_+(\nu+\Delta v)=1+\sum_{i=1}^{\infty}\sum_{l=0}^\infty c_{il}\,\rho^i\,(\Delta v)^l
\label{U+}
\end{equation}
under appropriate normalization of $\phi(u)$.
%%%%%%%%%%%%%%%%%%%%%%%%%%%%%%%%%%%%%%%%%%%%%%%%%%%%%%%%%%%%%%%%%%%%%%%%%%%%%%%%%%%%%%%%%%%%%%%%%%
\subsection{Perturbation Expansion}
To obtain the explicit semiclassical expansion for the spectrum $\nu_n$, we need:
\begin{enumerate}
\item \label{1}
to calculate several initial terms in  expansion (\ref{U+}),
\item \label{2}
to substitute  (\ref{U+}) into (\ref{reqs}) and to expand the expression in $\langle \ldots\rangle$
in powers of $\Delta v$,
\item
to perform averaging in (\ref{reqs}) by use of (\ref{av}),
\item \label{4}
to expand then $\arg \langle \ldots\rangle$ in (\ref{reqs}) in powers of $\rho$.
\end{enumerate}
Steps (\ref{2})-(\ref{4}) are straightforward and well suitable for computer calculations, below we describe 
only the step (\ref{1}).

Let us write down the formal Neumann series solving  equation (\ref{BS1})
in the class of the generalized functions:
\begin{eqnarray}
\phi(u)&=& \phi^{(0)}(u) +\phi^{(1)}(u)+O(\rho^2), \label{psi}\\
 \phi^{(0)}(u)&=&\frac{\pi}{\rmi \nu}[\delta(u-\nu)-\delta(u+\nu)], \nonumber \\
 \phi^{(1)}(u)&=&\frac{\rho\, u}{\rmi} \bigg[ \frac{1}{u^2-\nu^2} -8 \frac{1}{(u^2-\nu^2)^2}
+16 \frac{1-\nu^2}{(u^2-\nu^2)^3}\bigg],\label{psi1}
\end{eqnarray}
principal values are implied for the singular terms in (\ref{psi1}). 
Substitution of (\ref{psi}) into (\ref{g}), (\ref{a}),
(\ref{eqU+}) yields:
\begin{eqnarray*}
g(u)&=& g^{(0)}(u) + g^{(1)}(u) +O(\rho^2),\\
a&= &a^{(0)} +a^{(1)}+O(\rho^2),\nonumber\\
{\bf{U}}(u)&=&{\bf{U}}^{(0)}(u) + {\bf{U}}^{(1)}(u) +O(\rho^2), \nonumber
\end{eqnarray*}
where
\begin{eqnarray*}
g^{(0)}(u)&=& \frac{1}{u^2-\nu^2},\quad
a^{(0)}=-\frac{1}{2},\quad
{\bf{U}}^{(0)}(u)=1,\\
a^{(1)}&=&-\frac{\rho}{2\pi \nu^2}\bigg[-2+\nu^2+
\frac{-2-2\nu^2+\nu^4}{2\nu}\ln\bigg(\frac{1-\nu}{1+\nu}\bigg)\bigg].
\end{eqnarray*}
We skip lengthy expressions for $g^{(1)}(u)$ and  ${\bf{U}}^{(1)}(u)$. Note, that all singular terms at
$u=\nu$ and at $u=-\nu$   cancel in ${\bf{U}}^{(1)}(u)$. The explicit expressions for the Taylor coefficients $c_{il}$ in 
(\ref{U+}) read as
\begin{eqnarray*}
c_{10}&=&-\frac{\rmi\nu}{2},\\
c_{11}&=& \frac{-6+10\nu^2+3\nu^4-3\nu^6}{3\pi \nu^3 (\nu^2-1)^2}-
\frac{2+\nu^4}{2\pi \nu^4}\ln\bigg(\frac{1-\nu}{1+\nu}\bigg)-\frac{\rmi}{2},\nonumber\\
c_{12}&=& \frac{-9+27\nu^2-37\nu^4+3\nu^6}{6\pi \nu^4 (\nu^2-1)^3}-
\frac{\nu^2-3}{4\pi \nu^5}\ln\bigg(\frac{1-\nu}{1+\nu}\bigg), \nonumber\\
{\rm Im}\,c_{20}&=&\nu \,a^{(1)}/\rho. \nonumber
\end{eqnarray*}
These constants are sufficient to obtain the equation determining $\nu_n$ to the third order in 
$\rho$:
\begin{eqnarray}
-\frac{\Upsilon(\nu_n)}{4}=\rho \pi \big(n-\frac{1}{4}\big)+   \label{spe}
\rho^2\frac{5-6\nu_n^2-9\nu_n^4+6\nu_n^6}{12\nu_n^3(1-\nu_n^2)}\\
+ 
\rho^3 \bigg[\frac{30-62\nu_n^2+54\nu_n^4-21\nu_n^6+18\nu_n^8-3\nu_n^{10}}{6\pi\nu_n^5(\nu_n^2-1)^2}   \nonumber\\
+
\frac{10-4\nu_n^2+6\nu_n^4+4\nu_n^6-\nu_n^8}{4\pi\nu_n^6}\ln\bigg(\frac{1-\nu_n}{1+\nu_n}\bigg)\bigg]+O(\rho^4). \nonumber
\end{eqnarray}
In  variables $\theta_n$ and $\lambda$ related with $\nu_n$ and $\rho$ by
 $$
\nu_n=\tanh\theta_n, \qquad\rho=\frac{\lambda}{4 \cosh^2 \theta_n},
$$
equation (\ref{spe}) takes form (\ref{2M})-(\ref{S2}).
%%%%%%%%%%%%%%%%%%%%%%%%%%%%%%%%%%%%%%%%%%%%%%%%%%%%%%%%%%%%%%%
\section{Integral of the four-particle matrix element } \label{integ}
It was shown in section \ref{MCC}, that the local third-order multi-quark corrections to the meson mass 
can be expressed in terms of the integral of the four-particle matrix element
\begin{equation} \label{Wm}
\fl W(\beta_1,\beta_2)= 
\int_{-\infty}^\infty \rmd {\rm x} \int_{0}^\infty \rmd{\rm y} \lim_{\beta_1'\to \beta_1\atop\beta_2'\to \beta_2}
\langle\beta_2',\beta_1'| \sigma( {\rm x}, {\rm y})
(1- {\mathcal P}_0- {\mathcal P}_2)\sigma(0,0)
|\beta_1,\beta_2\rangle 
\end{equation}
over the Euclidean half-plane, see equations (\ref{M3low}), (\ref{M3c1}).  %in the limit $(\beta_1+\beta_2)/2\to\infty$. 
In this section we extract from $W(\beta_1,\beta_2)$ the most important "irreducible" part 
$W_{irr}(\beta_1,\beta_2)$, and transform
it to the form suitable for numerical calculations. 

It is straightforward to rewrite formula (\ref{E3S}), giving the local third-order semiclassical 
correction to the meson energy 
$\delta_3 E_n(P)$, in terms of  $W(\beta_1,\beta_2)$: 
\begin{eqnarray}
\label{E3}
\delta_3 E_n(P)=-\Bigg(\frac{h \bar{\sigma}}{m^2}\Bigg)^3 
\frac{4\,m^6}{E_n(P)\, (M_n^2-4 m^2 )} \frac{W(\beta_1,\beta_2)}{\bar{\sigma}^2}.
\end{eqnarray}
Here 
\begin{eqnarray*}
P=m\,(\sinh \beta_1+\sinh \beta_2), \\
E_n=m\,(\cosh \beta_1+\cosh \beta_2)
\end{eqnarray*} 
are the meson momentum and energy, and $\beta_1$ and $\beta_2$ are the rapidities of 
two quarks (forming the meson) at their collision, $M_n=(E_n^2-P^2)^{1/2}$ is the meson mass. 

The matrix element in the integrand in (\ref{Wm}) can be expanded by use of the Wick rule \cite{FZWard03}:
\begin{eqnarray}\label{Wi}
&& G(\beta_2,\beta_1|\beta_1,\beta_2) \equiv \lim_{\beta_1'\to \beta_1,\atop\beta_2'\to \beta_2}
\langle\beta_2',\beta_1'| \sigma( {\rm x}, {\rm y})\sigma(0,0)|\beta_1,\beta_2\rangle = \\
&& \frac{G(\beta_1|\beta_1)G(\beta_2|\beta_2)}{G}-\frac{ G(\beta_1|\beta_2) G(\beta_2|\beta_1)}{G}
-\frac{1}{G}\Bigg(\frac{ G(\beta_1,\beta_2) }{E(\beta_1)E(\beta_2)}\Bigg)^2.  \nonumber
\end{eqnarray}
Here we follow the notations of \cite{FZWard03}
\begin{eqnarray*}
 x=({\rm x},{\rm y})=(  r \cos\theta,r \sin \theta ),  \\
G(r)= \langle 0|\sigma( x )\sigma(0)|0\rangle , \\
 G( r,\theta;  \beta_1,\beta_2 ) = \langle 0|\sigma( x )\sigma(0) | \beta_1,\beta_2 \rangle, \\
\langle \beta'| \sigma( x )\sigma(0)|\beta\rangle=2 \pi \delta(\beta'-\beta)+ G(r,\theta;\beta'|\beta) , \\
E(r,\theta;\beta) =\exp\left[\frac{\rmi m r}{2} \sinh(\beta+\rmi\theta)\right],
\end{eqnarray*}
and drop the explicit indication of position dependence for the matrix elements. The equality 
$$
\langle \beta_2,\beta_1 |\sigma(x)\sigma(0)|0\rangle  = -\frac{\langle 0 |\sigma(x)\sigma(0)|\beta_1,\beta_2\rangle}
{[E(r,\theta;\beta_1)E(r,\theta;\beta_2)]^2}
$$
has been used in deriving (\ref{Wi}).
Explicit expressions for the matrix elements $G(r)$, $ G(r,\theta;  \beta_1,\beta_2 )$, 
 $G(r,\theta;\beta'|\beta)$ in terms of 
the solutions $\varphi(r)$, $\chi(r)$ of the Sinh-Gordon equation 
and associated Lax functions $\Psi_+(r,\theta;\beta)$, $\Psi_-(r,\theta;\beta)$
are known from \cite{FZWard03}.

Proceeding to polar coordinates $r, \theta$ in integral (\ref{Wm}), one can easily show, that 
\begin{equation}\label{trans}
\fl \int_0^\pi \rmd\theta \,G(r,\theta;\beta_2 ,\beta_1 |\beta_1  ,\beta_2 ) =
\int_0^\pi \rmd\theta \,G(r,\theta;\beta_2+ \beta ,\beta_1+\beta |\beta_1+ \beta,\beta_2+\beta ) 
\end{equation}
for arbitrary real $\beta$. Proof of (\ref{trans}) is based on the relations 
\begin{eqnarray} 
G(r,\theta;\beta_2,\beta_1|\beta_1,\beta_2) = 
G(r,0;\beta_2+\rmi \theta,\beta_1+\rmi \theta|\beta_1+\rmi \theta,\beta_2+\rmi \theta),\\
G(\beta_2 +\rmi \pi ,\beta_1 +\rmi \pi|\beta_1 +\rmi \pi,\beta_2+\rmi \pi)= 
G(\beta_2,\beta_1|\beta_1,\beta_2) ,  \label{mon}
\end{eqnarray}
which follow from (\ref{Wi}) and the similar properties of functions $G(\beta_1, \beta_2)$, 
$G(\beta_1|\beta_2)$, see \cite{FZ06}. 
Equality (\ref{trans}) means, that the integral in its left-hand side is Lorentz invariant. 

Unfortunately, the integral 
\[
 \int_{0}^\infty r \rmd r \int_{0}^\pi \rmd\theta \, G(r,\theta;\beta_2,\beta_1|\beta_1,\beta_2)
\]
diverges at large $r$. It becomes convergent after substraction of the "reducible part", see (\ref{Wm}):
\begin{eqnarray} \label{sub}
 G(r,\theta;\beta_2,\beta_1|\beta_1,\beta_2)  \to G(r,\theta;\beta_2,\beta_1|\beta_1,\beta_2) \\- 
 \lim_{\beta_1'\to \beta_1\atop\beta_2'\to \beta_2} \nonumber
\langle\beta_2'\beta_1'| \sigma(r,\theta)({\mathcal P}_0+ {\mathcal P}_2)\sigma(0,0)|\beta_1,\beta_2\rangle.
\end{eqnarray}
However, the second term in the right-hand side here does not satisfy the monodromy property like (\ref{mon}). 
This means, that the local multi-quark correction (\ref{E3}) to the meson energy is not Lorentz invariant by itself.
We hope that the Lorentz invariance form of $\delta_3  E_n (P)$ 
will be restored in the third order in $h$ after picking up all 
the  terms contributing to it, as it happens \cite{FZ06} for the second order term $\delta_2  E_n (P)$.   

At the moment it is helpful to extract the "Lorentz invariant" term from the reducible part in (\ref{sub}).
Namely, we shall subdivide it as follows:
\[
\fl \lim_{\beta_1'\to \beta_1\atop \beta_2'\to \beta_2} 
\langle\beta_2',\beta_1'| \sigma(r,\theta)({\mathcal P}_0+ {\mathcal P}_2)\sigma(0,0)|\beta_1,\beta_2\rangle =  
\Delta G(r,\theta;\beta_2,\beta_1|\beta_1,\beta_2) + \delta G(r,\theta;\beta_2,\beta_1|\beta_1,\beta_2),
\]
where the first term satisfies the required monodromy property
\begin{equation*}
\fl \Delta G(r,\theta;\beta_2 + \rmi \pi ,\beta_1+ \rmi \pi|\beta_1 + \rmi \pi,\beta_2+ \rmi \pi)= 
\Delta G(r,\theta;\beta_2,\beta_1|\beta_1,\beta_2),
\end{equation*}
while the function $\delta G(r,\theta;\beta_2,\beta_1|\beta_1,\beta_2)$ does not satisfy such a  property, but 
the  integral
\[
\int_{0}^\infty r \rmd r \int_{0}^\pi \rmd\theta \, \delta G(r,\theta; \beta_2,\beta_1|\beta_1,\beta_2)
\]
converges at finite $\beta_1, \;\beta_2$, 
and vanishes in the infinite momentum limit $(\beta_1+\beta_2)/2 \to +\infty$.

Note, that function $\Delta G(\beta_2,\beta_1|\beta_1,\beta_2) $ is analogous to function $S(\beta|\beta)$
defined by Equation (5.13) in Page 20 of reference \cite{FZWard03}, 
whereas $\delta G(\beta_2,\beta_1|\beta_1,\beta_2) $ is 
analogous to the zig-zag diagram (b) in figure~3 in  Page 19.  

Let us obtain explicit expressions  for $\Delta G(r,\theta; \beta_2  ,\beta_1|\beta_1,\beta_2  )$.

1. Vacuum sector.
\begin{eqnarray} \label{r0} 
\langle\beta_2,\beta_1| \sigma(r,\theta){\mathcal P}_0\sigma(0,0)|\beta_1,\beta_2\rangle= \\
 \bar{\sigma}^2 \tanh^2\frac{\beta_1-\beta_2}{2} \, 
\exp\{-{\rmi} r m [ \sinh(\beta_1+{\rmi} \theta) + \sinh(\beta_2+{\rmi}\theta)] \} = \nonumber\\
 2\, \bar{\sigma}^2 \tanh^2\frac{\beta_1-\beta_2}{2} \,  
\cos\{ r m [ \sinh(\beta_1+{\rmi}\theta) + \sinh(\beta_2+{\rmi}\theta)] \} - \nonumber \\
\bar{\sigma}^2 \tanh^2\frac{\beta_1-\beta_2}{2} \, 
\exp\{{\rmi}r m [ \sinh(\beta_1+{\rmi}\theta) + \sinh(\beta_2+{\rmi}\theta)] \}. \nonumber
\end{eqnarray}
The first and the second terms in the right-hand side of (\ref{r0}) 
should be assigned to $\Delta G(\beta_2,\beta_1|\beta_1,\beta_2) $ 
and to $\delta G(\beta_2,\beta_1|\beta_1,\beta_2) $, respectively. 

2. Two-quark sector. 
\begin{eqnarray} \nonumber
\fl \langle\beta_2',\beta_1'| \sigma(r,\theta){\mathcal P}_2\,\sigma(0,0)|\beta_1,\beta_2\rangle= 
\int_{-\infty}^\infty \frac{\rmd\eta_1 \, \rmd \eta_2}{(2 \pi)^2} 
\langle\beta_2',\beta_1'| \sigma(r,\theta)| \eta_2 ,\eta_1 \rangle
\langle \eta_1 ,\eta_2|\sigma(0,0)|\beta_1,\beta_2\rangle. 
\end{eqnarray}
This can be splitted into five  diagrams: 
\begin{eqnarray} \label{dgr}
\raisebox{-.8em} {\includegraphics[width=6em]{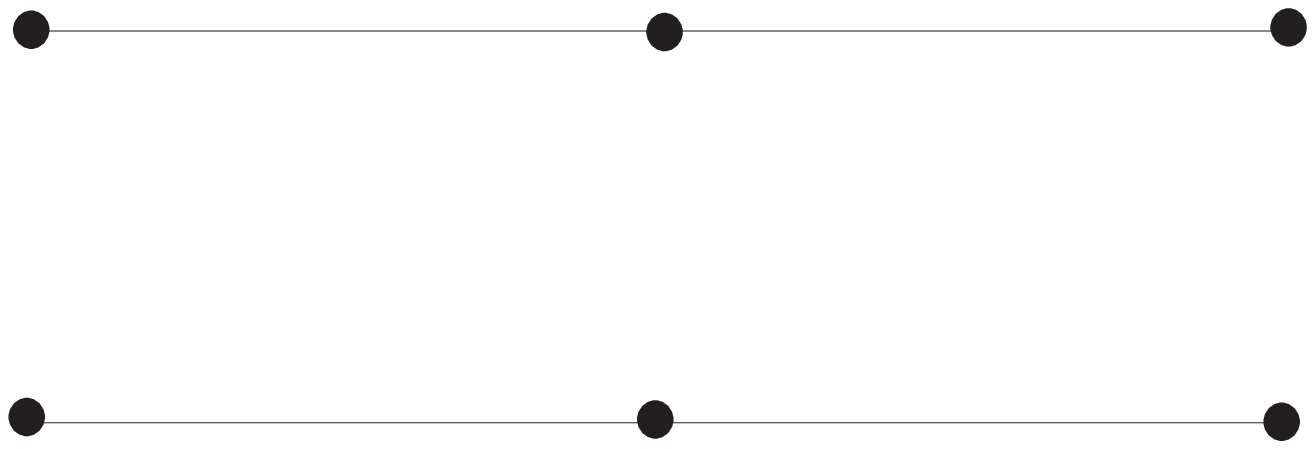}} \;\;\; +\;\;\;
\raisebox{-.8em} {\includegraphics[width=6em]{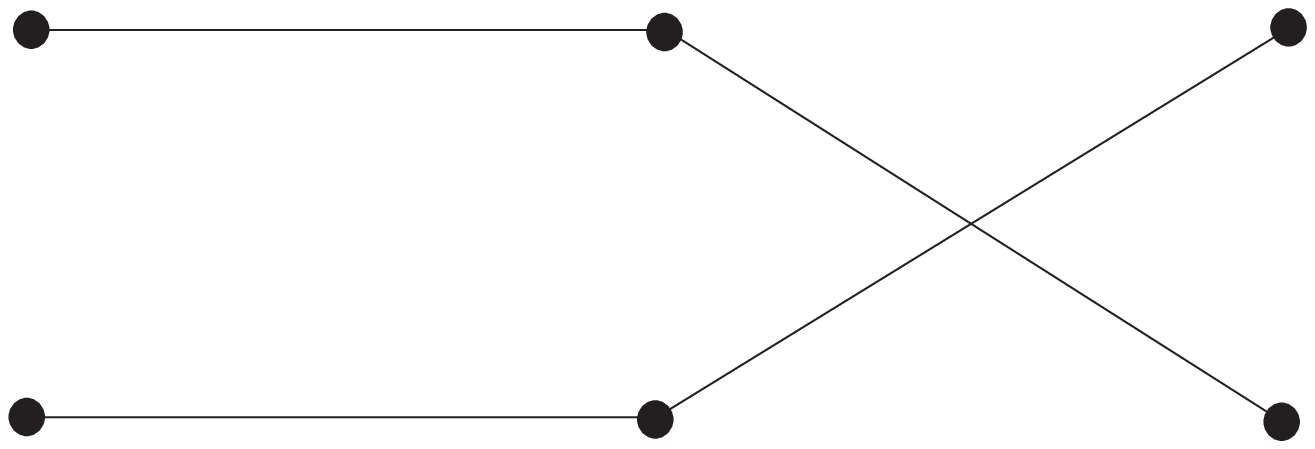}}  \;\;\; +\;\;\;
\raisebox{-.8em} {\includegraphics[width=6em]{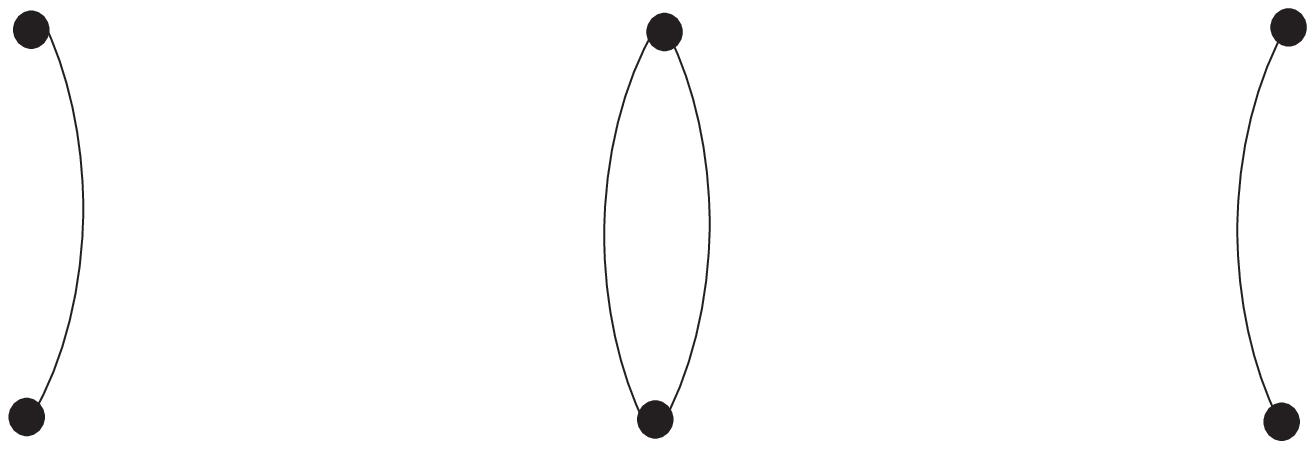}} \;\; \;\\+\;\;\;
\raisebox{-.8em} {\includegraphics[width=6em]{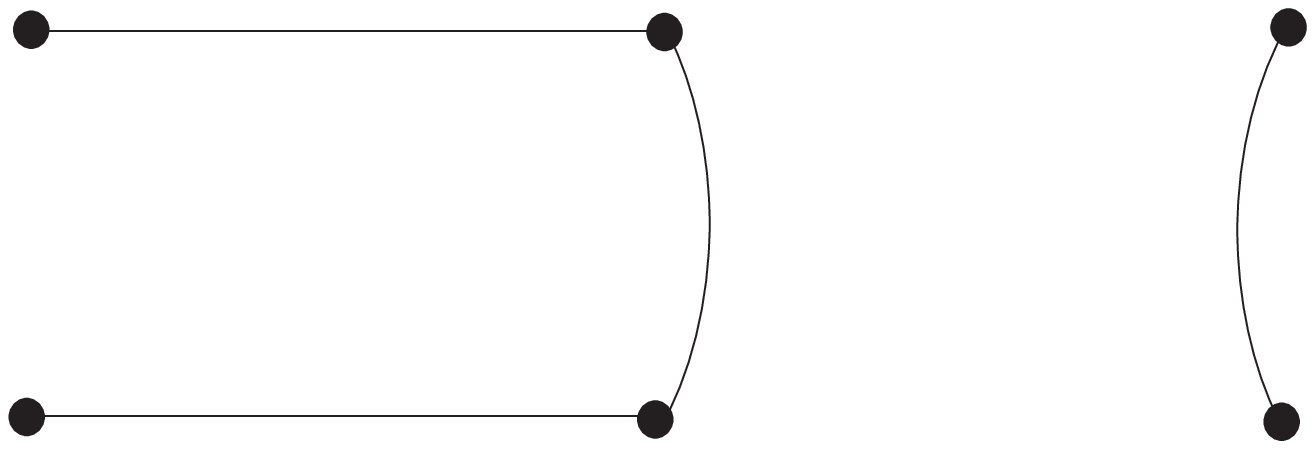}} \;\;\; +\;\;\;
\raisebox{-.8em} {\includegraphics[width=6em]{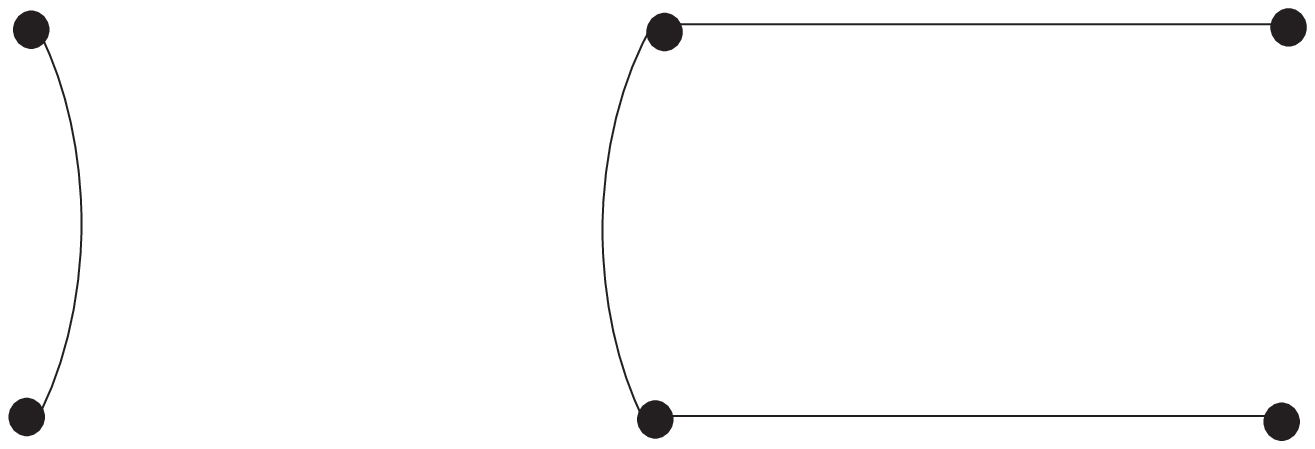}} \nonumber
\end{eqnarray}
Contribution of two former diagrams into  $\Delta G(\beta_2,\beta_1|\beta_1,\beta_2) $  are: 
\begin{equation}
\raisebox{-.8em} {\includegraphics[width=6em]{./d1a.eps}}\;
\to\;\frac{1}{\bar{\sigma}^{2}}S(\beta_1 |\beta_1) S( \beta_2 |\beta_2), 
\end{equation}
\begin{equation}
\raisebox{-.8em} {\includegraphics[width=6em]{./d2a.eps}}\;
\to\;- \frac{1}{\bar{\sigma}^{2}}[R (\beta_1 |\beta_2)]^2,
\end{equation} 
where 
\begin{eqnarray} \label{R}
\fl R(\beta_1|\beta_2) = \bar{\sigma}^2 \exp[{-\rmi m r(\sinh \beta_1+\sinh \beta_2)/2}]
\int_{-\infty}^\infty \frac{\rmd\eta}{2\pi}\exp[{\rmi mr\sinh\eta}]\\
\fl\cdot \coth\frac{\eta-\beta_1}{2} \coth\frac{\eta-\beta_2}{2}+
2\bar{\sigma}^2 \coth\frac{\beta_1-\beta_2}{2} \, \sin[mr(\sinh \beta_1-\sinh \beta_2)/2]\nonumber\\\fl+ 
\bar{\sigma}^2 \exp[{\rmi m r(\sinh \beta_1+\sinh \beta_2)/2}]
\int_{-\infty}^\infty \frac{\rmd\eta'}{2\pi}\exp[{\rmi mr\sinh\eta'}] 
\tanh\frac{\eta'-\beta_1}{2}\tanh\frac{\eta'-\beta_2}{2},\nonumber \\
\fl S(\beta_1|\beta_1)=\lim_{\beta_2\to \beta_1}R(\beta_1|\beta_2). \nonumber
\end{eqnarray}
Note, that $0<{\rm Im}\beta_j <\pi$ for $j=1,2$  is supposed in (\ref{R}). It is easy to verify, that:
\[
R(\beta_1+\rmi \pi|\beta_2+\rmi\pi)=  R(\beta_1|\beta_2) 
\]

The third diagram in (\ref{dgr}) is proportional to the function $ f^{(2)}(t) $, which determines 
the well-known large-distance 
asymptotics of the Ising correlation function \cite{McCoy76}. Its contribution to 
$\Delta G(\beta_2,\beta_1|\beta_1,\beta_2) $ is 
\begin{eqnarray*}
\fl\raisebox{-.8em} {\includegraphics[width=6em]{./d3a.eps}} \;\;
\to- 2 \bar{\sigma}^2 \cos\{ r m [ \sinh(\beta_1+\rmi \theta) + \sinh(\beta_2+\rmi \theta)] \} 
\tanh^2\frac{\beta_1-\beta_2}{2} f^{(2)}(m r),\\
 f^{(2)}(t)=-\frac{1}{\pi^2}\{[ K_1^2(t) -K_0^2(t)]t^2-t K_0(t)\, K_1(t) +\frac{1}{2} K_0^2(t)\},
\end{eqnarray*}
where $K_0(t),\, K_1(t)$ are the MacDonald's functions.

The forth and the fifth diagrams are equal to one another and can be written as:
\begin{eqnarray} 
\fl\raisebox{-.8em} {\includegraphics[width=6em]{./d5a.eps}}% \;\; 
=- \rmi \bar{\sigma}^2 \tanh \frac{\beta_1-\beta_2}{2} 
\Big\{1+ \exp\left[-\rmi mr[ \sinh(\beta_1+i\theta) 
+ \sinh(\beta_2+\rmi\theta)]\right] \Big\} \nonumber\\ 
\cdot{\mathcal T}_2 (r,\beta_1+\rmi \theta,\beta_2+\rmi\theta)
+\bar{\sigma}^2 \tanh^2  \frac{\beta_2-\beta_1}{2}
 \label{di45}\\ 
\fl +\rmi \,\bar{\sigma}^2 \tanh \frac{\beta_1-\beta_2}{2}\,\,
\exp\{-\rmi mr[ \sinh(\beta_1+\rmi\theta) + \sinh(\beta_2+\rmi\theta)]\} 
{\mathcal A}_4(r,\beta_1+\rmi \theta,\beta_2+\rmi\theta)\nonumber \\
 + \rmi \bar{\sigma}^2 \tanh \frac{\beta_1-\beta_2}{2}\,\,
{\mathcal T}_2 (r,\beta_1+ i \theta,\beta_2 +\rmi\theta) ,\nonumber 
\end{eqnarray}
where
\begin{equation*}
{\mathcal T}_2 (r,\beta_1,\beta_2) = {\mathcal B}_2 (r,\beta_1,\beta_2) +{\mathcal {U}}_2 (r,\beta_1,\beta_2)+
{\mathcal V}_2 (r,\beta_1,\beta_2)+{\mathcal A}_4 (r,\beta_1,\beta_2),
\end{equation*}
and
\begin{eqnarray} \nonumber\label{F2}
\fl{\mathcal B}_2 (r,\beta_1,\beta_2)=-\rmi\int_{-\infty}^\infty
\frac{\rmd\eta_1\,\rmd\eta_2}{(2\pi)^2}e^{\rmi mr( \sinh \eta_1 + \sinh \eta_2) }
\coth\frac{\eta_1-\beta_1}{2} \coth\frac{\eta_2-\beta_2}{2} \tanh\frac{\eta_1-\eta_2}{2}, \\
\fl{\mathcal U}_2 (r,\beta_1,\beta_2)=-\int_{-\infty}^\infty\frac{\rmd\eta_2}{2\pi}
e^{\rmi mr( \sinh \beta_1 + \sinh \eta_2) }
 \coth\frac{\eta_2-\beta_2}{2} \tanh\frac{\beta_1-\eta_2}{2}, \nonumber\\
\fl{\mathcal V}_2 (r,\beta_1,\beta_2)=-\int_{-\infty}^\infty\frac{\rmd\eta_1}{2\pi}
e^{\rmi mr( \sinh \eta_1 + \sinh \beta_2) }
 \coth\frac{\eta_1-\beta_1}{2} \tanh\frac{\eta_1-\beta_2}{2}, \nonumber\\
\fl{\mathcal A}_4 (r,\beta_1,\beta_2)=-\rmi e^{\rmi mr(\sinh \beta_1 + \sinh \beta_2)}\, 
\int_{-\infty}^\infty\frac{\rmd\eta_1\,\rmd\eta_2}{(2\pi)^2}e^{\rmi mr( \sinh \eta_1 + \sinh \eta_2) } \nonumber\\
\lo\cdot \tanh\frac{\eta_1-\beta_1}{2}\,\, \tanh\frac{\eta_2-\beta_2}{2}\,\, \tanh\frac{\eta_1-\eta_2}{2}.\nonumber
\end{eqnarray}
Here we again suppose $0<{\rm Im}\beta_j <\pi$ for $j=1,2$.

Note, that
$$
{\mathcal T}_2 (r,\beta_1+\rmi\pi,\beta_2+\rmi \pi)=
\exp[{-\rmi mr(\sinh \beta_1 + \sinh \beta_2)}]\,{\mathcal T}_2 (r,\beta_1,\beta_2).
$$
The two former terms in (\ref{di45}) contribute to 
$\Delta G(\beta_2,\beta_1|\beta_1,\beta_2) $, while all the rest terms
in  (\ref{di45}) should be assigned to $\delta G(\beta_2,\beta_1|\beta_1,\beta_2) $. 
Thus, the irreducible part of the two-particle matrix element takes the form:
\begin{eqnarray} \label{Gir}
\fl G_{irr}(r,\theta;\beta_2,\beta_1|\beta_1,\beta_2) \equiv G(r,\theta;\beta_2,\beta_1|\beta_1,\beta_2)-
\Delta G(r,\theta;\beta_2,\beta_1|\beta_1,\beta_2)\\ \nonumber
\fl =\left[\frac{G(\beta_1|\beta_1)G(\beta_2|\beta_2)}{G}-\frac{S( \beta_1|\beta_1) 
S( \beta_2|\beta_2)}{\bar{\sigma}^2}\right]- \nonumber
\left[\frac{ G(\beta_1|\beta_2) G(\beta_2|\beta_1)}{G}- \frac{  [R (\beta_1 |\beta_2)]^2 }{\bar{\sigma}^{2}}\right]\\-
\left[\frac{1}{G}\left(\frac{ G(\beta_1,\beta_2) }{E(\beta_1)E(\beta_2)}\right)^2 + 
C_2(\beta_1,\beta_2) \right],\nonumber
\end{eqnarray}
where
\begin{eqnarray*}
\fl C_2(\beta_1,\beta_2)\equiv  C_2(r,\theta;\beta_1,\beta_2)=2  \bar{\sigma}^2 \tanh^2  \frac{\beta_2-\beta_1}{2}   \\
\fl +
2\, \bar{\sigma}^2 \tanh^2\frac{\beta_1-\beta_2}{2} \,  
\cos\{ r m [ \sinh(\beta_1+\rmi \theta) + \sinh(\beta_2+\rmi \theta)] \} [1-f^{(2)}(mr)]\\\fl-
 2\rmi \bar{\sigma}^2 \tanh \frac{\beta_1-\beta_2}{2}\,
\Big\{1+ \exp\{-\rmi mr[ \sinh(\beta_1+\rmi\theta) + \sinh(\beta_2+\rmi\theta)]\} \Big\}\\
\lo\cdot{\mathcal T}_2 (r,\beta_1+\rmi \theta,\beta_2+\rmi\theta).  \nonumber 
\end{eqnarray*}
Integration of (\ref{Gir}) in $r$ and $\theta$ gives the irreducible part of the factor $ W(\beta_1,\beta_2) $:
\begin{equation} \label{Wirr}
W_{irr}(\beta_1,\beta_2) = \int_0^\infty r  \rmd r\int_0^\pi \rmd\theta \,G_{irr}(r,\theta;\beta_2,\beta_1|\beta_1,\beta_2).
\end{equation}
Integral in $r$ here is convergent for small enough $|\beta_1-\beta_2|$, and 
\[W_{irr}(\beta_1,\beta_2)=W_{irr}(\beta_1+\beta,\beta_2+\beta)\]
for arbitrary $\beta$.

On the other hand, the integral 
\[
\int_0^\infty r  \rmd r\int_0^\pi \rmd\theta \,\delta G(r,\theta;\beta_2,\beta_1|\beta_1,\beta_2)
\]
converges and vanishes in the infinite momentum frame.

So, the local multi-quark contribution to the third-order meson  mass correction takes the form: 
\begin{eqnarray}
\label{M3}
 \delta_3 M_n^2=-\Bigg(\frac{h \bar{\sigma}}{m^2}\Bigg)^3 
\frac{8\,m^6}{M_n^2-4 m^2 } \frac{ W_{irr}(\beta_1,\beta_2) }{\bar{\sigma}^2},
\end{eqnarray}
with $W_{irr}(\beta_1,\beta_2)$ given by (\ref{Wirr}). Three other third-order contributions to $\delta M_n^2$ 
come from the two-fermion Bethe-Salpeter equation in the infinite momentum frame (\ref{BSi}), and 
from the quark mass and string tension renormalizations.
\section*{Referencses}

\end{document}